\documentclass[lettersize,journal]{IEEEtran}

\usepackage[table]{xcolor}
\colorlet{shadecolor}{gray!15}
\colorlet{shadeblue}{blue!15}

\usepackage{amsmath,amsfonts}
\usepackage{algorithmic}
\usepackage{algorithm}
\usepackage{array}
\usepackage[caption=false,font=normalsize,labelfont=sf,textfont=sf]{subfig}
\usepackage{textcomp}
\usepackage{stfloats}
\usepackage{url}
\usepackage{verbatim}
\usepackage{graphicx}
\usepackage{cite}
\usepackage{enumitem}
\usepackage{hyperref}
\usepackage{makecell}
\usepackage{etoolbox}
\robustify\bfseries
\usepackage{amssymb}
\usepackage{tabularx}
\usepackage{wasysym}
\usepackage{color,soul}
\usepackage{balance}
\usepackage{booktabs}
\usepackage{multirow}
\usepackage{siunitx}
\usepackage{arydshln}
\usepackage{multirow}

\def\H{{\mathsf H}}
\def\T{{\mathsf T}}
\def\CC{{\mathbb C}}
\def\RR{{\mathbb R}}
\usepackage{caption}
\usepackage{amssymb}%
\usepackage{pifont}%

\usepackage{tikz}

\newcommand{\argmin}{\operatornamewithlimits{argmin}}

\begin{document}

\title{Cross-Talk Speech Reduction, by Separation, for Separation}

\author{Zhong-Qiu Wang,~\IEEEmembership{Member,~IEEE,} and Samuele Cornell,~\IEEEmembership{Member,~IEEE}
\thanks{Manuscript received on April 30, 2026. 
\textit{(Corresponding author: Zhong-Qiu Wang).}
}
\thanks{
Z.-Q. Wang is with the Department of Computer Science and Engineering, Southern University of Science and Technology, Shenzhen 518055, China (e-mail: wang.zhongqiu41@gmail.com / wangzq3@sustech.edu.cn).}
\thanks{S. Cornell is with the Language Technologies Institute, Carnegie Mellon University, Pittsburgh, PA 15213, USA (e-mail: cornellsamuele@gmail.com).}
}

\maketitle

\begin{abstract}
In conversational speech separation and recognition tasks, close-talk microphones are typically attached to each speaker during training data collection to capture near-field, close-talk mixture signals, in addition to using far-field microphones to record far-field mixture signals.
Each such close-talk mixture exhibits a reasonably high
energy level for the wearer and could intuitively serve as weak supervision for training far-field speech separation models directly on real-recorded far-field signals.
However, they are not sufficiently clean for this purpose, as they often contain strong cross-talk speech from other speakers in addition to background noise.
To address this, we propose \emph{cross-talk reduction} (CTR), a task aiming to isolate the wearer's speech from each close-talk mixture, and a novel method called CTRnet, which can be trained directly on real-recorded pairs of close-talk and far-field mixtures to accomplish CTR.
Building on CTRnet, we further propose pseudo-label based far-field speech separation (PuLSS), which uses CTRnet's estimated clean speech as pseudo-labels to train models for separating far-field mixtures.
A key advantage of the proposed framework is that both CTRnet and PuLSS
can be trained on real-recorded data from the target domain, addressing the generalization gap commonly observed when models are trained exclusively on simulated data.
On the CHiME-6 dataset, our framework achieves state-of-the-art ASR performance under both oracle and estimated speaker diarization, surpassing all CHiME-\{7,8\} challenge submissions.
To our knowledge, it is the first neural speech separation method that substantially outperforms guided source separation on real conversational ``speech-in-the-wild'' data.

\end{abstract}

\section{Introduction}

\begin{figure}
  \begin{center}
  \includegraphics[width=8.5cm]{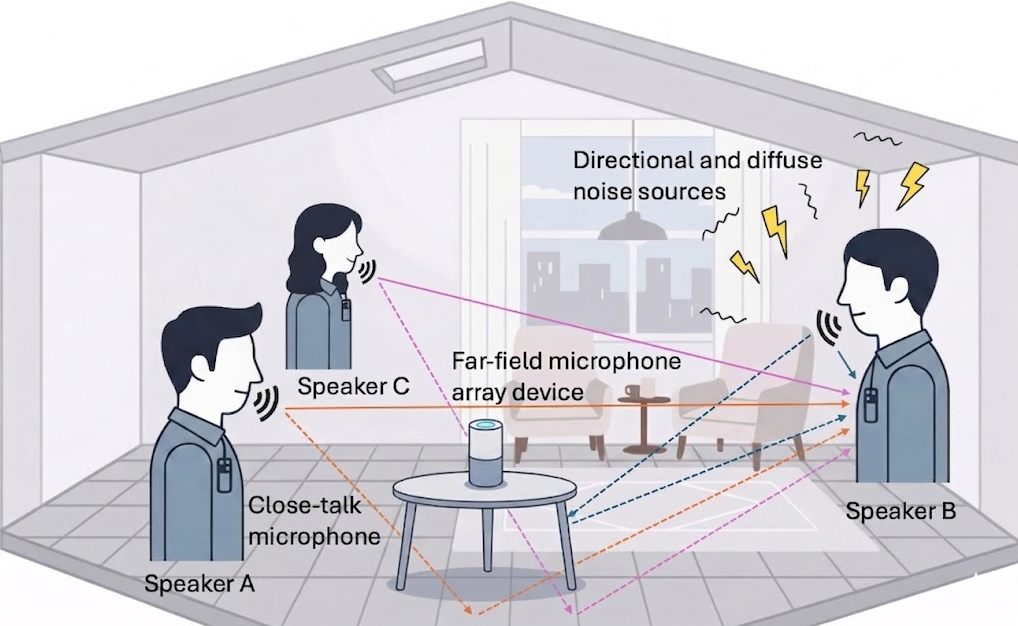}
  \end{center}
  \vspace{-0.3cm}
  \caption{
  Typical setup for collecting training data in conversational scenarios, where a microphone array is placed at far field to record far-field mixtures.
  For annotation purposes,
  a standard practice
  is to have each speaker wear a close-talk microphone
  to record close-talk mixtures so that data annotation (e.g., word transcriptions and annotating speaker-activity timestamps) can be facilitated.
  At deployment, only the far-field array is available for inference.
  }\label{physical_model_figure}
  \vspace{-0.5cm}
\end{figure}

\IEEEPARstart{I}{n} many pattern analysis and machine intelligence applications, sensors, while recording target signals, inevitably capture concurrent non-target signals (i.e. interference) within the same environment, resulting in a mixture of target and non-target signals \cite{comon2010handbook}.
The non-target signals often pose
difficulties for the perception and understanding of the target signals.
A prominent example, in audio and acoustic signal processing, is conversational speech separation, \textit{a.k.a.}, the cocktail party problem \cite{McDermott2009,WDLreview,Araki2025}.
In realistic acoustic environments and conversational scenarios (see Fig.~\ref{physical_model_figure}), speech signals are corrupted by ambient noise, room reverberation, and interference from other speakers. The goal of speech separation is to isolate the speakers of interest from the mixture signals, allowing downstream tasks such as automatic speech recognition (ASR) to perform well \cite{Haeb-Umbach2020,Haeb-Umbach2025}.
Depending on the application, the recording device may range from a single microphone to a microphone array or a set of distributed arrays, with signals typically captured in far-field conditions, where the speaker-to-microphone distance is large relative to the array aperture.

In the past decade, major advances have been made in speech separation, driven primarily by the adoption of deep learning \cite{WDLreview,Araki2025}. The current dominant approach relies on supervised learning using large-scale synthetic datasets.
To generate these, clean anechoic speech signals from different speakers are first convolved with room impulse responses (RIRs) to simulate reverberant conditions.
The resulting signals are then mixed with background noises at varying energy levels to create noisy-reverberant multi-speaker mixtures.
Finally, deep neural networks (DNN) are trained on these mixture–target pairs to learn to predict the clean target speech from noisy-reverberant mixtures \cite{WDLreview}.
The clean speech paired with the mixture provides a perfect supervision \textit{at the sample level}, which allows to fully leverage the power of supervised deep learning.
However, although showing impressive performance on simulated mixtures \cite{Wang2023TFGridNet}, such fully-supervised, synthetically-trained models often exhibit limited generalizability to real-recorded mixtures \cite{Zhang2021ClosingGap, Subakan2022RealM, abramovski2025summary, Cornell2025CHiME78Analysis, masuyama2026end}, as current simulation techniques cannot simulate mixtures that are sufficiently realistic, resulting in a persistent domain-mismatch problem between simulated model training and real-world deployment.

A natural solution
is to train speech separation models directly on real-recorded mixtures in the target domain, thereby mitigating the domain mismatch problem.
However, 
for real-recorded mixtures, the individual source signals are not available (unlike in the simulated case, where clean sources are available by simulation).
As a result, there lacks \textit{high-quality sample-level supervision signals} for training supervised separation models on real-recorded mixtures.

When collecting far-field conversational data for ASR or speaker diarization, it is a standard practice\footnote{This recording setup is employed in most far-field conversational speech datasets such as AMI \cite{McCowan2006}, AliMeeting \cite{Yu2022M2MeT}, CHiME \cite{Barker2018CHiME5} and MISP \cite{Wang2023MISP}.} to simultaneously record each speaker through a microphone placed near their mouth, such as a lapel microphone (see Fig.~\ref{physical_model_figure}).
We refer to 
each such recording 
as a \emph{close-talk mixture},
inside which each wearer's own speech,
namely \emph{close-talk speech},
has a much higher energy level than in any far-field mixture, making close-talk mixtures a natural candidate for supervision when training models to separate real-recorded far-field mixtures.
However, close-talk mixtures are not clean.
They often contain significant cross-talk from other speakers and ambient noise, albeit at lower energy levels than in far-field mixtures.
Consequently, close-talk mixtures,
which are also mixtures of multiple sound sources, generally cannot be used directly as pseudo-labels for training separation models on real-recorded far-field mixtures.
These observations motivate our study of cross-talk reduction: separating each speaker's close-talk speech from its close-talk mixture. Once separated, these signals can serve as high-quality pseudo-labels and enable supervised training of separation models directly on real-recorded far-field mixtures, thereby mitigating domain mismatches.

A preliminary version \cite{Wang2023CTRnet} of this work has been published in the IJCAI conference, but only addresses \textit{cross-talk reduction by separation} on close-talk mixtures.
This paper improves \textit{cross-talk reduction by separation} on close-talk mixtures and extends it \textit{for separation} on far-field mixtures.
Specifically, in \cite{Wang2023CTRnet}, we have made the following contributions:
\begin{itemize}[leftmargin=*,noitemsep,topsep=0pt]
\item We introduce a task named cross-talk reduction and propose to formulate it as a blind deconvolution problem, which requires estimating both the close-talk speech of each speaker and the relative transfer functions (RTF) relating the close-talk speech to reverberant speech at other microphones.
\item We propose a solution, CTRnet, which is unsupervised in nature and can be directly trained on pairs of real-recorded close-talk and far-field mixtures.
\item We extend this latter to weakly-supervised CTRnet, where speaker-activity timestamps are leveraged as a weak supervision to improve the training of unsupervised CTRnet.
\end{itemize}
Building upon the conference paper \cite{Wang2023CTRnet}, this paper further makes the following contributions:
\begin{itemize}[leftmargin=*,noitemsep,topsep=0pt]
\item We propose semi-supervised CTRnet, which is trained by combining supervised training if the input mixture is simulated and weakly-supervised training if it is real-recorded.
\item We develop a novel mechanism to include noise modeling in CTRnet to deal with ambient noises.
\item We introduce a novel mechanism to reduce the reverberation of close-talk speech in CTRnet.
\item We propose a pseudo-label approach named PuLSS for far-field speech separation, where close-talk speech estimated by CTRnet is leveraged to derive pseudo-labels for training supervised separation models on real-recorded far-field mixtures (to predict the pseudo-labels).
In this way, we can train separation models directly on real-recorded signals in the target domain, potentially realizing better separation.
\end{itemize}
\noindent
The proposed PuLSS system obtains state-of-the-art conversational ASR performance on the real-recorded, notoriously-difficult CHiME-6 dataset \cite{Barker2018CHiME5}, representing a practical step toward solving the cocktail party problem in real-world conditions.
A sound demo is provided in the link below\footnote{See \url{https://zqwang7.github.io/demos/CTRnet_journal_demo/index.html}.}.

\section{Related Work}\label{related_work_description}

This paper is related to existing work mainly in two aspects.

\subsection{Neural Speech Separation for ASR
}\label{ssec:frontend_separation_related}

Although much progress has been made in supervised neural speech separation, the success of using it as a frontend processing for robust ASR in realistic conversational conditions is limited \cite{Haeb-Umbach2020, Cornell2025CHiME78Analysis, Haeb-Umbach2025}, largely due to the aforementioned domain mismatch problem between simulated training and real-recorded test conditions.
Many studies have observed that the predicted target speech by the trained models often exhibits severe speech distortion, which is very detrimental to ASR.

To improve ASR performance, hybrid methods, which combine
supervised DNN-based separation with signal processing based linear filtering, have been proposed.
One representative approach is to leverage DNN-separated signals to compute signal statistics (e.g., spatial covariance matrices) for
linear beamforming, and the beamforming results are then used for ASR.
This approach is often effective \cite{Heymann2015, erdogan2016improved, masuyama2026end}, as beamforming is linearly constrained and thus inherently limits speech distortion \cite{iwamoto2022bad}.
However, the DNNs still remain sensitive to domain mismatches \cite{masuyama2026end}, 
and linear filtering itself typically cannot produce sufficient separation in realistic acoustic scenarios, especially when the number of available microphones is limited.
Fine-tuning strategies, which jointly fine-tune neural speech separation frontends with a backend ASR model in an end-to-end fashion \cite{masuyama2026end, kanda2023vararray}, can produce better ASR performance, but joint ASR fine-tuning often degrades the quality of the separated signals themselves \cite{masuyama2026end}.

To improve speech separation itself in realistic conditions, several un- and weakly-supervised approaches, which can be trained directly on target-domain un- or weakly-labeled mixtures, have been proposed.
Mixture invariant training \cite{wisdom2020unsupervised} and its variants based on the mixture-of-mixtures concept \cite{zhang2021teacher, saijo2023self, han2024unsupervised} have shown potential \cite{sivaraman2022adapting}, but they inherently struggle from issues such as source permutation ambiguity, reliance on synthetically-mixed mixtures, lacking dereverberation capabilities, and over- and under-separation \cite{wisdom2021sparse}, making their applicability to real-world conversational scenarios challenging.
Another stream of research realizes unsupervised separation by exploiting multi-channel recordings. 
For example, UNSSOR \cite{Wang2023UNSSOR} and enhanced RAS \cite{saijo2024enhanced} impose mixture constraints among different microphone observations, and DNN-IVA \cite{saijo2022spatial} fuses classical blind source separation with DNNs.
However, they only demonstrated effectiveness under simulated conditions with overly simplified setup.
Their performance in realistic conversational scenarios is unclear. 

Practically, however, none of these previously proposed neural separation methods can produce a satisfactory performance in real-world conversational scenarios, due to challenges in dealing with, e.g., non-stationary ambient noises, time-varying number of speakers, sparse overlap among speakers, long-form recordings (which require producing consistent speaker output channel with no ambiguity), and low signal quality (due to, e.g., microphone failures and synchronization issues) which is very common in real-world deployment.
This has been sufficiently demonstrated in recent benchmarks.
For example, in the recent CHiME challenges \cite{Cornell2025CHiME78Analysis}, all systems relied on guided source separation (GSS) \cite{boeddeker2018front}, a signal processing algorithm which leverages speaker-activity timestamps and
complex-valued Gaussian mixture models to derive signal statistics for linear beamforming.
Neural approaches such as VarArray \cite{yoshioka2022vararray} and continuous source separation \cite{chen2020continuous, vinnikovnotsofar, von2024meeting} have shown promise, but remain less robust even within a single acoustic domain.
This is evidenced by their inferior performance compared to GSS-based approaches in the recent NOTSOFAR-1 challenge \cite{abramovski2025summary}.
Among the few exceptions are recently-proposed unsupervised full-rank spatial covariance models \cite{bando2023neural, bando2024neural, bando2025investigation}, which have demonstrated competitive and slightly-superior performance to GSS on CHiME-8 conversational data \cite{bando2025investigation}, albeit only with oracle diarization.

In contrast, our proposed PuLSS approach trains separation models directly on real-recorded far-field mixtures using pseudo-labels derived from close-talk mixture signals, thereby avoiding the domain mismatch problems that limit existing neural approaches.
To the best of our knowledge, PuLSS is the first neural separation method to significantly outperform GSS in real-recorded conversational scenarios, with both oracle and estimated diarization (see later Section \ref{comparison_with_existing_studies_description_oracle_diar} and \ref{comparison_with_existing_studies_description_estimated_diar}).

\subsection{Exploiting Close-Talk Mixtures for Speech Enhancement}

There are studies exploiting close-talk mixtures for far-field speech enhancement, a task similar to speech separation but dealing with a single target speaker.
ctPuLSE \cite{Wang2024ctPuLSE} first enhances real-recorded close-talk mixtures by using supervised models trained on simulated mixtures, and then uses the enhanced close-talk speech as pseudo-labels for training far-field speech enhancement models.
However, the supervised model used to derive pseudo-labels also suffers from domain mismatch problems when enhancing real-recorded close-talk mixtures.
SuperM2M \cite{Wang2024SuperM2M}, building upon M2M \cite{Wang2024M2M}, trains DNNs on far-field mixtures such that the DNN estimates can be linearly filtered to approximate the speech and noise components in close-talk mixtures. However, this technique assumes a single speaker and is not natively designed to leverage the fact that the close-talk mixture exhibits a higher signal-to-noise ratio (SNR) of the wearer.
SuPseudo \cite{Luo2025SuPseudo} and TLS \cite{Luo2025TLS} linearly filter unprocessed close-talk mixtures and use them as pseudo-labels for far-field speech enhancement.
This approach assumes that close-talk mixtures are sufficiently clean, which is not the case in most conversational scenarios.

Unlike these methods, our approach handles multi-speaker separation rather than single-speaker enhancement, does not assume clean close-talk mixtures, and avoids domain mismatch by training CTRnet directly on real-recorded mixtures rather than relying on supervised models trained on simulated mixtures to derive pseudo-labels.

\section{Physical Model and Objectives}\label{physical_model_and_objectives}

Suppose that we have a set of training mixtures recorded in a number of noisy-reverberant environments, each with a far-field microphone array with $P$ microphones and a maximum of $C$ speakers (each wearing a close-talk microphone near the mouth or on the lapel).
See Fig.~\ref{physical_model_figure} for an illustration.
The physical models for each close-talk mixture and each far-field mixture can be formulated, in the short-time Fourier transform (STFT) domain, as follows:
\begin{align}
Y_d(t,f) &= \sum\nolimits_{c=1}^C X_d(c,t,f) + V_d(t,f), \label{physical_model_ct} \\
Y_p(t,f) &= \sum\nolimits_{c=1}^C X_p(c,t,f) + V_p(t,f), \label{physical_model_ff}
\end{align}
where $c$ indexes $C$ speakers, $d$ indexes $C$ close-talk microphones (as each speaker wears a single close-talk microphone), $p$ indexes $P$ far-field microphones, $t$ indexes $T$ frames, and $f$ indexes $F$ frequency bins.
In Eq. (\ref{physical_model_ct}), $Y_d(t,f)$, $X_d(c,t,f)$, and $V_d(t,f)$ respectively denote the STFT coefficients of the mixture, reverberant speech (or speaker image) of speaker $c$, and reverberant noise signals captured by close-talk microphone $d$ at time $t$ and frequency $f$.
In this paper, when dropping the indices $c$, $t$ and $f$, we refer to the corresponding spectrograms.
In Eq. (\ref{physical_model_ff}), $Y_p$, $X_p(c)$ and $V_p$ respectively denote the STFT spectrograms of the mixture, reverberant speech of speaker $c$, and reverberant noise signals captured by far-field microphone $p$.
We denote the close-talk speech of each speaker $c$ as ``$X_d(c)$ with $d=c$'' or ``$X_{d(=c)}(c)$'', and denote the cross-talk speech of speaker $c$ at close-talk microphone $d$ as ``$X_d(c)$ with $d\neq c$''.

\begin{figure}[t]
  \begin{center}
  \includegraphics[width=0.5\textwidth]{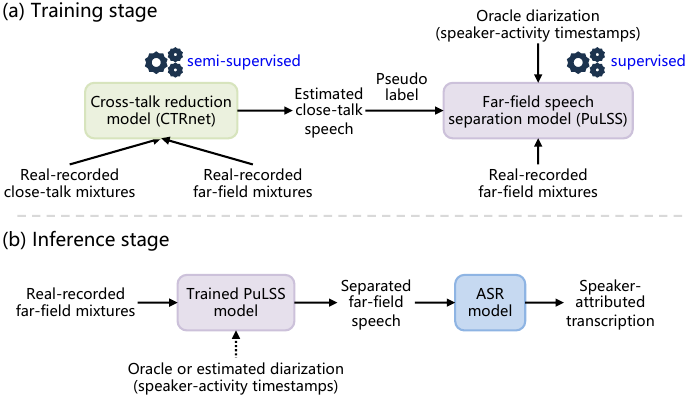}
  \end{center}
  \vspace{-0.3cm}
  \caption{
System overview.
(a)~\textbf{Training Stage}: CTRnet is trained in a semi-supervised manner on real-recorded pairs of close-talk and far-field  mixtures to estimate close-talk speech (see Section~\ref{ssec:semi_sup_ctrnet}). The estimate is then used as pseudo-labels for training PuLSS in a supervised fashion on real-recorded far-field mixtures (see Section~\ref{ssec:pulss_real_simu}).
In PuLSS, oracle speaker-activity timestamps are used in input features to resolve the speaker-permutation problem during training.
(b)~\textbf{Inference Stage}: the PuLSS model separates far-field mixtures and an ASR model transcribes the separated speech.
At inference time, either oracle speaker-activity timestamps or estimated ones by an external speaker diarization system can be used (see the dashed arrow).
}
\label{system_overview_figure}
\vspace{-0.3cm}
\end{figure}

With this formulation, we propose to study \textit{cross-talk reduction by speech separation}, aiming at reducing the cross-talk speech and noises in each close-talk mixture to estimate the close-talk speech, followed by \textit{cross-talk reduction for speech separation}, where the resulting estimated close-talk speech are leveraged to derive pseudo-labels for training supervised far-field speech separation models.
See Fig. \ref{system_overview_figure} for an overview.

A straightforward way to realize cross-talk reduction is to first simulate many pairs of close-talk mixtures and close-talk speech, and then train a supervised model to predict the close-talk speech based on the close-talk mixtures.
However, this supervised approach based on simulated data often exhibits limited generalizability to real-recorded mixtures, as discussed in the Introduction section.

We propose to train models for cross-talk reduction directly on target-domain, real-recorded pairs of close-talk and far-field mixtures, in an un-, weakly- or semi-supervised way, thereby improving the generalizability.
In the following sections, 
we propose CTRnet
in Section \ref{CTRnet_description} to estimate close-talk speech, followed by a model named PuLSS in \ref{PuLSS_description}, which leverages estimated close-talk speech to compute pseudo-labels for training supervised far-field speech separation models.

\section{Cross-Talk Reduction by Separation}\label{CTRnet_description}

We propose CTRnet, which can be trained via unsupervised and weakly-supervised learning on real-recorded pairs of close-talk and far-field mixtures to realize cross-talk reduction.
To avoid confusion, Table \ref{summary_hyperparam} lists the hyper-parameters of our models, and their default values or set of values to tune.

\begin{table*}[t]
\scriptsize
\centering
\sisetup{table-format=2.2,round-mode=places,round-precision=2,table-number-alignment = center,detect-weight=true,detect-inline-weight=math}
\caption{\textsc{List of Key Hyper-Parameters of CTRnet and PuLSS.}}
\vspace{-0.1cm}
\label{summary_hyperparam}
\setlength{\tabcolsep}{0.5pt}
\begin{tabular}{
ccccc
}
\toprule

Model & Symbols & Description & Introduced in & Values\\

\midrule

{\multirow{6}{*}{CTRnet}} & $I$ & Number of past taps in FCP filtering & Eq. (\ref{physical_model_ct_reformulate}) & $13$ \\
 & $J$ & Number of future taps in FCP filtering & Eq. (\ref{physical_model_ct_reformulate}) & $1$ \\
 & $\xi$ & Factor for flooring denominator in FCP & Eq. (\ref{FCPweight_ct}) & $0.01$ \\
 & $\beta$ & Weight of speaker-activity loss & Eq. (\ref{loss_MC+SA}) & $\{1.0,0.1\}$ \\
 & $\Delta$ & Prediction delay for modeling reverberation & Eq. (\ref{loss_MC_close_talk_one_addnoise_further_dereverb}) & $\{1,2,3,4\}$ \\
 & $\kappa_1$ & \makecell{Weight for supervised loss on simulated data in semi-supervised CTRnet} & Eq. (\ref{CTRnet_loss_simu_real}) & $1.0$ \\
\midrule
{\multirow{5}{*}{PuLSS}} & $L$ & Number of filter taps to compute pseudo-labels & Eq. (\ref{pseudo_direct_RIR}) & $2$ \\
& $E$ & Maximum hypothesized time delay for synchronization & Eq. (\ref{estimate_K}) & $9$ \\ 
& $A$ & Number of filters taps on each side for $\mathcal{L}_{\text{CTE}}$ loss & Eq. (\ref{CTE_loss}) & $1$ \\
 & $\delta$ & Weight of loss on close-talk estimates & Eq. (\ref{total_PuLSS_real_loss}) & $20$ \\
 & $\kappa_2$ & \makecell{Weight for supervised loss on simulated data when training PuLSS with both simu and real data} & Eq. (\ref{PuLSS_loss_simu_real}) & $1.0$ \\
\midrule
\multirow{2}{*}{Both CTRnet \& PuLSS} & $\alpha$ & Magnitude compression factor & Eq. (\ref{loss_F}) & $\{1.0,0.3\}$\\
& $\theta$ & Factor in weighted sampling & Eq. (\ref{weighted_sampling}) & $\{5,10,20,40,80\}$\\
\bottomrule
\end{tabular}
\vspace{-0.5cm}
\end{table*}

\subsection{Formulating CTR as Blind Deconvolution}\label{blind_deconv_description}

Due to the short distance from each speaker to its close-talk microphone, the close-talk speech can be viewed as the dry sound source signal with a small time-delay\footnote{Although some reverberation exists in close-talk speech, it is much weaker than the direct-path signal while the speaker is talking.}.
In this case, linearly filtering close-talk speech can largely reproduce its cross-talk speech at the close-talk microphones of the other speakers, as well as the reverberant speech of the speaker at far-field microphones.
With this understanding, we can reformulate the physical models in Eq. (\ref{physical_model_ct}) and (\ref{physical_model_ff}) in the following as (\ref{physical_model_ct_reformulate}) and (\ref{physical_model_ff_reformulate}), and formulate CTR as a blind deconvolution problem in (\ref{ideal_loss}).

In detail, let $Z(d) = X_{d(=c)}(c)$ denote the close-talk speech at close-talk microphone $d$, we can formulate Eq. (\ref{physical_model_ct}) as
\begin{align}
&Y_d(t,f) = Z(d,t,f) + \sum\limits_{c=1,c\neq d}^C X_d(c,t,f) + V_d(t,f) \nonumber \\
&=Z(d,t,f) + \sum\limits_{c=1,c\neq d}^C \mathbf{g}_{d}(c,f)^{\H}\ \widetilde{\mathbf{Z}}(c,t,f) + V_d'(t,f), \label{physical_model_ct_reformulate}
\end{align}
where $\widetilde{\mathbf{Z}}(c,t,f)=[Z(c,t-I,f),\dots,Z(c,t,f),\dots,Z(c,t+J,f)]^\T \in \CC^{I+1+J}$ stacks the complex STFT coefficients of a time window of $I+1+J$ time-frequency (T-F) units within frequency bin $f$.
Here, $I$ and $J$ respectively denote the number of past and future filter taps, and $\mathbf{g}_{d}(c,f) \in \CC^{I+1+J}$ is a linear filter.
In Eq. (\ref{physical_model_ct_reformulate}), we have $X_d(c,t,f) \approx \mathbf{g}_{d}(c,f)^{\H}\ \widetilde{\mathbf{Z}}(c,t,f)$, which, following narrowband linear approximation \cite{Talmon2009CTF,Gannot2017}, approximates the cross-talk speech of speaker $c$ captured by close-talk microphone $d$ (i.e., $X_d(c)$ with $d\neq c$) as a linear convolution between the close-talk speech of speaker $c$ (i.e., $Z(c)$) and an RTF relating the close-talk speech of speaker $c$ to close-talk microphone $d$ (i.e., $\mathbf{g}_{d}(c,f)$ where $d\neq c$).
In Eq. (\ref{physical_model_ct_reformulate}), $V_d'$ absorbs the modeling error of linear approximation.
Similarly, for far-field mixtures, we can reformulate Eq.(\ref{physical_model_ff}) as
\begin{align}
Y_p(t,f) &= \sum\limits_{c=1}^C \mathbf{g}_{p}(c,f)^{\H}\ \widetilde{\mathbf{Z}}(c,t,f) + V_p'(t,f), \label{physical_model_ff_reformulate}
\end{align}
with $\mathbf{g}_{p}(c,f)$ denoting the RTF from speaker $c$ to far-field microphone $p$ and $V_p'$ absorbing the modeling error.

With the physical models in Eq. (\ref{physical_model_ct_reformulate}) and (\ref{physical_model_ff_reformulate}) and assuming $V'$ (including ambient noises, and modeling errors incurred by linear approximation) being small, we can realize cross-talk reduction by solving, e.g., the minimization problem below:
\begin{align}\label{ideal_loss}
&\underset{\mathbf{g}_{\cdot}(\cdot,\cdot),Z(\cdot,\cdot,\cdot)}{\argmin}
\Big( \nonumber \\ 
&\sum\limits_{d=1}^C \sum\limits_{t,f} \Big| Y_d(t,f) - Z(d,t,f) - \sum\limits_{\substack{c=1, c\neq d}}^C {\mathbf{g}}_{d}(c,f)^{\H}\ \widetilde{{\mathbf{Z}}}(c,t,f) \Big|^2 \nonumber \\
&\quad\,\,\,\,\,+\sum\limits_{p=1}^{P} \sum\limits_{t,f} \Big| Y_p(t,f) - \sum\limits_{c=1}^C {\mathbf{g}}_p(c,f)^{\H}\ \widetilde{{\mathbf{Z}}}(c,t,f) \Big|^2 \Big),
\end{align}
which aims at finding the linear filters and close-talk speech signals that are, in a least-square sense, most consistent with the physical models in (\ref{physical_model_ct_reformulate}) and (\ref{physical_model_ff_reformulate}).
This problem is an embodiment of the blind deconvolution problem \cite{Levin2011} in pattern analysis and machine intelligence.
It is difficult to solve as both the linear filters and close-talk speech signals are unknown and need to be estimated, but only the summation of their linear-convolutional results (i.e., the close-talk and far-field mixtures) are observed.
To deal with this, our preliminary conference paper \cite{Wang2023CTRnet} proposes to solve this problem via unsupervised deep learning, yielding unsupervised CTRnet, which is described next.

\begin{figure*}
  \begin{center}
  \includegraphics[width=11.25cm]{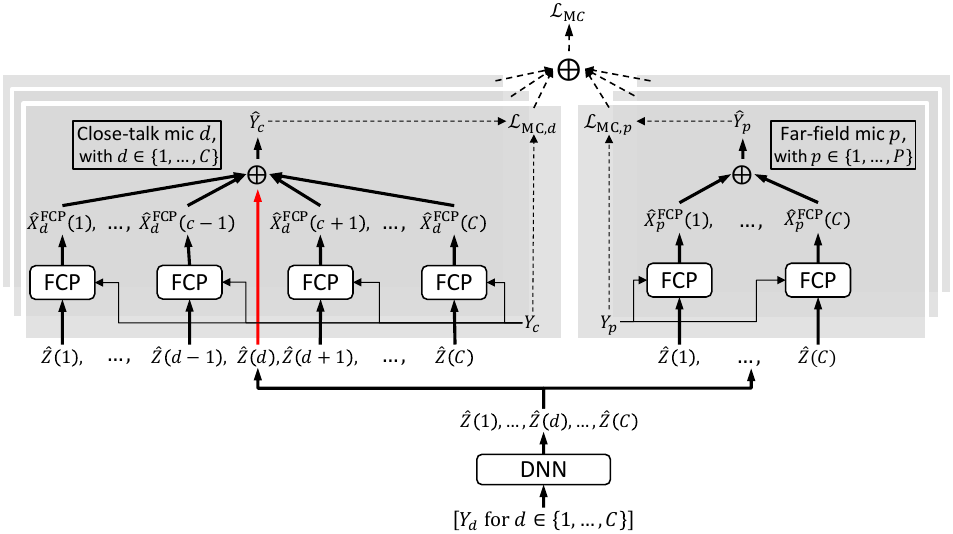}
  \end{center}
  \vspace{-0.3cm}
  \caption{
  Illustration of unsupervised CTRnet.
  Best viewed in color.
  }\label{unsupervised_CTRnet_figure}
  \vspace{-0.4cm}
\end{figure*}

\subsection{Unsupervised CTRnet}\label{unsupervised_ctrnet}

Fig. \ref{unsupervised_CTRnet_figure} illustrates unsupervised CTRnet, which trains a DNN, using all the $C$ close-talk mixtures as input, to produce an estimate $\hat{Z}(d)\in\CC^{T\times F}$ at each close-talk microphone $d$.
Differently from the supervised setup, for real-recorded close-talk mixtures, we do not have oracle target speech to directly penalize the estimates $\{\hat{Z}(d)\}_{d=1}^C$.
In unsupervised CTRnet, we propose to penalize the estimates by checking to what extent the estimates satisfy the physical models hypothesized in Eq. (\ref{physical_model_ct_reformulate}) and (\ref{physical_model_ff_reformulate}), thereby promoting the estimates to approximate the corresponding close-talk speech.
We realize this penalization by considering each of the close-talk and far-field mixtures as a constraint to the estimates, and define a mixture-constraint (MC) loss, which follows the objective in Eq. (\ref{ideal_loss}):
\begin{align}\label{loss_MC}
\mathcal{L}_{\text{MC}} = \sum\nolimits_{d=1}^C \mathcal{L}_{\text{MC},d} + \sum\nolimits_{p=1}^P  \mathcal{L}_{\text{MC},p},
\end{align}
where $\mathcal{L}_{\text{MC},d}$ is the MC loss at close-talk microphone $d$ and $\mathcal{L}_{\text{MC},p}$ at far-field microphone $p$.
Following the physical model in (\ref{physical_model_ct_reformulate}) and the first term in (\ref{ideal_loss}), at close-talk microphone $d$ we define $\mathcal{L}_{\text{MC},d}$ as
\begin{align}
&\mathcal{L}_{\text{MC},d} = \sum_{t,f} \mathcal{F} \Big( Y_{d}(t,f), \hat{Y}_{d}(t,f) \Big) \nonumber \\
&= \sum\limits_{t,f} \mathcal{F} \Big( Y_{d}(t,f), \hat{Z}(d,t,f) + \sum_{c=1,c\neq d}^C \hat{X}_{d}^{\text{FCP}}(c,t,f) \Big) \nonumber \\
&= \sum\limits_{t,f} \mathcal{F} \Big( Y_{d}(t,f), \hat{Z}(d,t,f) + \sum_{c=1,c\neq d}^C \hat{\mathbf{g}}_{d}(c,f)^{\H}\ \widetilde{\hat{\mathbf{Z}}}(c,t,f) \Big). \label{loss_MC_close_talk_one}
\end{align}
where, following the definitions of $\widetilde{\mathbf{Z}}(c,t,f)$ and $\mathbf{g}_{d}(c,f)$ in Eq. (\ref{physical_model_ct_reformulate}), $\widetilde{\hat{\mathbf{Z}}}(c,t,f)=[\hat{Z}(c,t-I,f),\dots,\hat{Z}(c,t,f),\dots,\hat{Z}(c,t+J,f)]^\T \in \CC^{I+1+J}$ stacks a time window of $I+1+J$ T-F units within frequency bin $f$ and $\hat{\mathbf{g}}_{d}(c,f) \in \CC^{I+1+J}$ is an estimated linear filter used for filtering the estimated close-talk speech of the other speakers to approximate their cross-talk speech at close-talk microphone $d$ (i.e., $\hat{X}_{d}^{\text{FCP}}(c,t,f)=\hat{\mathbf{g}}_{d}(c,f)^{\H}\ \widetilde{\hat{\mathbf{Z}}}(c,t,f)$).
We estimate the linear filter (i.e., $\hat{\mathbf{g}}_{d}(c,f)$) via the forward convolutive prediction (FCP) algorithm \cite{Wang2021FCPjournal}, which will be described later in Eq. (\ref{fcp_proj_mixture}).
Finally, we compute a loss between the reconstructed mixture $\hat{Y}_d$ and the observed mixture $Y_d$, by using a loss function $\mathcal{F}(\cdot, \cdot)$, which, following \cite{Wang2023TFGridNet}, computes an absolute loss on the estimated real, imaginary and magnitude components:
\begin{align}
&\mathcal{F} \Big( Y_d(t,f), \hat{Y}_d(t,f) \Big) = \frac{\mathcal{G} \Big( Y_d(t,f), \hat{Y}_d(t,f) \Big)}{\sum\nolimits_{t',f'} \big| Y_d(t',f') \big|^\alpha}, \label{loss_F} \\
&\mathcal{G} \Big( Y_d(t,f), \hat{Y}_d(t,f) \Big) = \Big| |Y_d(t,f)|^\alpha  - |\hat{Y}_d(t,f)|^\alpha \Big| \nonumber \\
&+\Big| |Y_d(t,f)|^\alpha \cos(\angle Y_d(t,f)) - |\hat{Y}_d(t,f)|^\alpha \cos(\angle \hat{Y}_d(t,f)) \Big| \nonumber \\
&+\Big| |Y_d(t,f)|^\alpha \sin(\angle Y_d(t,f)) - |\hat{Y}_d(t,f)|^\alpha \sin(\angle \hat{Y}_d(t,f)) \Big|. \label{loss_G}
\end{align}
In Eq. (\ref{loss_F}), the denominator is a normalization term balancing the losses at different microphones.
Different from the conference paper \cite{Wang2023CTRnet}, this paper introduces a tunable magnitude compression factor $\alpha$ to the loss function, following \cite{Wisdom2018MixtureConsistency}.

Similarly, following the physical model in (\ref{physical_model_ff_reformulate}) and the second term in (\ref{ideal_loss}), at each far-field microphone $p$ we define $\mathcal{L}_{\text{MC},p}$ as
\begin{align}
\mathcal{L}_{\text{MC},p} &= \sum_{t,f} \mathcal{F} \Big( Y_p(t,f), \hat{Y}_p(t,f) \Big) \nonumber \\
&= \sum_{t,f} \mathcal{F} \Big( Y_p(t,f), \sum_{c=1}^C  \hat{X}_p^{\text{FCP}}(c,t,f) \Big) \nonumber \\
&= \sum_{t,f} \mathcal{F} \Big( Y_p(t,f), \sum_{c=1}^C \hat{\mathbf{g}}_p(c,f)^{\H}\ \widetilde{\hat{\mathbf{Z}}}(c,t,f) \Big), \label{loss_MC_far_field}
\end{align}
where we linearly filter the DNN estimate $\hat{Z}(c)$ for each speaker $c$ using $\hat{\mathbf{g}}_p(c,f)$ so that their summation can approximate the observed far-field mixture.

The linear filters, $\hat{\mathbf{g}}_{d}(c,f)$ in Eq. (\ref{loss_MC_close_talk_one}) and $\hat{\mathbf{g}}_p(c,f)$ in (\ref{loss_MC_far_field}), are estimated via FCP by solving the following problem \cite{Wang2021FCPjournal}:
\begin{align}\label{fcp_proj_mixture}
&\hat{\mathbf{g}}_m(c,f) = \underset{\mathbf{g}_m(c,f)}{\text{argmin}}
\sum\limits_t \frac{\Big| Y_m(t,f)-\mathbf{g}_m(c,f)^{\H}\ \widetilde{\hat{\mathbf{Z}}}(c,t,f) \Big|^2}{\lambda_m(t,f)},
\end{align}
where $m$ indexes the $C$ close-talk and $P$ far-field microphones, and $\lambda$ is a weighting term defined, following \cite{Wang2021FCPjournal}, as
\begin{align}\label{FCPweight_ct}
\lambda_m(t,f) = \xi\times \max(|Y_m|^2) + |Y_m(t,f)|^2,
\end{align}
with $\xi$ flooring the weighting term and $\max(\cdot)$ extracting the maximum value of a power spectrogram.
In this study, we propose another way to compute the weighting term:
\begin{align}
\lambda_m(t,f) &= \xi\times \operatorname{quantile}\big(\Omega,90\big) + |Y_m(t,f)|^2, \label{FCPweight_ct_2}
\end{align}
where $\Omega = \{\max(|Y_m(t,\cdot)|^2)\}_{t=1}^T$ consists of the maximum energy of the T-F units within each frame, and $\operatorname{quantile}\big(\Omega,90\big)$ extracts the $90$-th percentile.
We find that this strategy can more effectively deal with the case when the input mixture has clicking sounds with sudden power bursts (e.g., from microphones moving around or inadvertently touched), which is quite common in real-recorded conversational signals.
Notice that the optimization problem in Eq. (\ref{fcp_proj_mixture}) is a linear regression problem.
It has a closed-form solution which can be readily computed.
We then plug the solution into Eq. (\ref{loss_MC_close_talk_one}) and (\ref{loss_MC_far_field}) to compute the MC losses, and train the network.

\subsection{Weakly-Supervised CTRnet}

\begin{figure}
  \begin{center}
  \includegraphics[width=0.8\columnwidth]{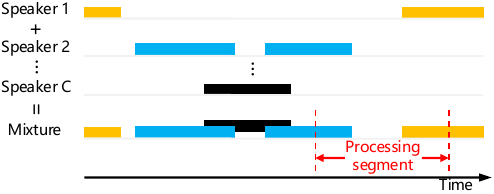}
  \end{center}
  \vspace{-0.3cm}
  \caption{Illustration of sparse and time-varying speaker overlap.
  Each colored block indicates that the corresponding speaker is active. The bracket denotes a fixed-length processing segment used for both training and inference (see Section~\ref{block_split_description}), within which the number of active speakers may vary.}
  \label{sparse_overlap_figure}
  \vspace{-0.3cm}
\end{figure}

Unsupervised CTRnet is trained on fixed-length mixture segments, assuming at maximum $C$ active speakers
to separate within each segment.
This assumption however does not always hold as the number of speakers is different for different training segments.
See Fig. \ref{sparse_overlap_figure} for an example.
When the active number of speakers is smaller than $C$, unsupervised CTRnet tends to over-separate the speakers (e.g., split a speaker to multiple outputs, as doing this would always result in a smaller MC loss); and when the active number of speakers is larger than $C$, unsupervised CTRnet would under-separate the speakers (i.e., cannot sufficiently separate the speakers).
These behaviors are similar to what we observe in unsupervised clustering algorithms such as Kmeans when the hypothesized number of clusters differs from the actual number. %

To address this issue, we assume that, for each speaker $c$, a speaker-activity timestamp label $d(c)\in \{0,1\}^N$ denoting whether each speaker $c$ is active at each sample (assuming the signal is $N$-sample long) is provided. 
As mentioned in the Introduction section, in practical setups for collecting real-recorded conversational data, such labels are routinely 
annotated,
and can even be obtained in a semi-automatic way by using additional on-speaker throat microphones \cite{vinnikovnotsofar}. 

The timestamps are leveraged as a weak-supervision to improve the training of unsupervised CTRnet, as they can provide the information of the exact number of active speakers at each sample.
They are leveraged to mask DNN estimates $\hat{Z}$ before computing the FCP filters and $\mathcal{L}_{\text{MC}}$ loss:
\begin{align}\label{muting}
\hat{Z}(c,t,f):=\hat{Z}(c,t,f) \times D(c,t),
\end{align}
where $D(c,t)\in \{0, 1\}$, defined based on $d(c)$, is one if the STFT window corresponding to frame $t$ contains any active speech samples of speaker $c$ and is zero otherwise.
We name this technique \textit{frame muting}.

After using frame muting in Eq. (\ref{muting}), the $\mathcal{L}_{\text{MC}}$ loss only penalizes DNN predictions in non-silent ranges marked by the speaker-activity timestamps.
However, the predictions in silent ranges are no longer penalized.
To deal with this, we introduce a speaker-activity (SA) loss $\mathcal{L}_{\text{SA}}$ to push the DNN estimate towards zero:
\begin{align}\label{loss_SA}
\mathcal{L}_{\text{SA}} = \sum\limits_{c=1}^C  
\frac{\sum_{t,f}\big| \hat{Z}(c,t,f) \big|^\alpha \times \big( 1-D(c,t) \big)}
{\sum_{t,f} \big| Y_{d(=c)}(t,f) \big|^\alpha},
\end{align}
where the denominator $\sum_{t,f} |Y_{d(=c)}(t,f)|^\alpha$ is the compressed energy of the close-talk mixture signal of speaker $c$, serving as a normalization term consistent with Eq. (\ref{loss_F}).
We combine it with $\mathcal{L}_{\text{MC}}$ in (\ref{loss_MC}) for model training:
\begin{align}\label{loss_MC+SA}
\mathcal{L}_{\text{MC+SA}} = \mathcal{L}_{\text{MC}} + \beta \times \mathcal{L}_{\text{SA}},
\end{align}
where $\beta \in \RR_{>0}$ is a tunable weighting term.

\subsection{Semi-Supervised CTRnet}\label{ssec:semi_sup_ctrnet}

CTRnet can be trained directly on real-recorded pairs of close-talk and far-field mixtures.
However, real-recorded training mixtures are often scarce, as collecting them is labor-intensive.
In this context, we propose a semi-supervised learning algorithm that trains the same CTRnet model on both real-recorded and simulated mixtures.
When the input mixture is real-recorded, we use the weakly-supervised loss in Eq. (\ref{loss_MC+SA}) for training, and when the input mixture is simulated, we use a supervised loss defined as
\begin{align}
\mathcal{L}_{\text{sup,speech}}^\text{CTRnet} = \sum_{c=1}^{C} \frac{ \sum_{t,f} \mathcal{G} \Big( \hat{Z}(c,t,f), X_{d (=c)}(c,t,f) \Big) }{ \sum_{t,f}| Y_{d (=c)}(t,f)|^\alpha },\label{loss_supervised_speech}
\end{align}
where $\mathcal{G}(\cdot,\cdot)$ is defined in Eq. (\ref{loss_G}), and $X_{d (=c)}(c)$ and $Y_{d (=c)}$ respectively denote the close-talk speech and close-talk mixture of speaker $c$.
The overall loss is defined as
\begin{equation}
\mathcal{L}_\text{sup,MC+SA}^{\text{CTRnet}} = \left\{
\begin{aligned}
&\kappa_1 \times \mathcal{L}_{\text{sup}}^\text{CTRnet}, \text{if input mixture is simulated}, \\
&\mathcal{L}_{\text{MC+SA}}, \text{if input mixture is real-recorded},
\end{aligned}
\right.
\label{CTRnet_loss_simu_real}
\end{equation}
where $\kappa_1\in \RR_{\geq 0}$ is a tunable weighting term, and $\mathcal{L}_{\text{sup}}^\text{CTRnet}$ summates all the supervised losses consisting of $\mathcal{L}_{\text{sup,speech}}^\text{CTRnet}$ in Eq. (\ref{loss_supervised_speech}) and, optionally, $\mathcal{L}_{\text{sup,noise}}^\text{CTRnet}$ defined later in (\ref{loss_supervised_noise}).

\subsection{Noise Modeling}

The CTRnet models presented so far assume that ambient noises (i.e., $V$) are weak, and do not model noises.
This would result in estimated close-talk speech signals inevitably containing
noises, as the estimated signals are trained, via the MC losses, to reconstruct observed mixture signals, which are usually noisy.
To alleviate this issue, we propose to model noises by training the DNN model to additionally predict the noise signal at each close-talk microphone. 
In other words, the DNN model is trained to output $2\times C$ signals, with the first $C$ outputs estimating close-talk speech and the rest estimating the noise signals at the close-talk microphones.
When the input mixture is simulated, we penalize the noise estimates using the following supervised loss:
\begin{align}
\mathcal{L}_{\text{sup,noise}}^\text{CTRnet} = \sum_{d=1}^{C} \frac{ \sum_{t,f} \mathcal{G} \Big( \hat{V}_d(t,f), V_d(t,f) \Big) }{ \sum_{t,f}| Y_d(t,f)|^\alpha},\label{loss_supervised_noise}
\end{align}
where $\hat{V}_d$ denotes the DNN-estimated noise STFT spectrogram at close-talk microphone $d$, and $V_d$ the corresponding oracle (available only for simulated mixtures).
When the input mixture is real-recorded, we penalize them via MC losses.
In detail, we first average the noise estimates and consider the average as a source in addition to the $C$ speakers:
\begin{align}
\hat{Z}(C+1)=\frac{1}{C}\sum_{d=1}^C \hat{V}_d,\label{noise_averaging}
\end{align}
and modify $\mathcal{L}_{\text{MC},d}$ defined in Eq. (\ref{loss_MC_close_talk_one}) and $\mathcal{L}_{\text{MC},p}$ in (\ref{loss_MC_far_field}) to include the noise estimate $\hat{Z}(C+1)$:
\begin{align}
&\mathcal{L}_{\text{MC},d} = \sum\limits_{t,f} \mathcal{F} \Big( Y_{d}(t,f), \nonumber \\
&\quad\quad\quad\quad\quad\,\,\hat{Z}(d,t,f) + \sum_{c=1,c\neq d}^{C+1} \hat{\mathbf{g}}_{d}(c,f)^{\H}\ \widetilde{\hat{\mathbf{Z}}}(c,t,f) \Big), \label{loss_MC_close_talk_one_addnoise} \\
&\mathcal{L}_{\text{MC},p} = \sum_{t,f} \mathcal{F} \Big( Y_p(t,f), \sum_{c=1}^{C+1} \hat{\mathbf{g}}_p(c,f)^{\H}\ \widetilde{\hat{\mathbf{Z}}}(c,t,f) \Big). \label{loss_MC_far_field_addnoise}
\end{align}
An alternative is to randomly choose a noise estimate as $\hat{Z}(C+1)$ for each training example:
\begin{align}
\hat{Z}(C+1)=\operatorname{RandomChoice}\big(\{\hat{V}_d\}_{d=1}^C\big).\label{noise_random}
\end{align}

For the additional source, we do not apply frame muting, since noise-activity timestamp is not available. We just assume that the noise is always active.

\subsection{Reverb Modeling and Dereverb of Close-Talk Speech}

Reverberation exists in close-talk speech.
In previous sections, it is assumed much weaker than the direct-path signal and negligible.
In this subsection, we explicitly model it in close-talk speech in order to reduce it.
In detail, we modify the loss function in Eq. (\ref{loss_MC_close_talk_one_addnoise}) to
\begin{align}
&\mathcal{L}_{\text{MC},d} = \sum\limits_{t,f} \mathcal{F} \Big( Y_{d}(t,f), \nonumber \\
&\hat{Z}(d,t,f) + \hat{\mathbf{h}}_{d}(f)^{\H}\ \overline{\hat{\mathbf{Z}}}(d,t,f) + \sum_{c=1,c\neq d}^{C+1} \hat{\mathbf{g}}_{d}(c,f)^{\H}\ \widetilde{\hat{\mathbf{Z}}}(c,t,f) \Big), \label{loss_MC_close_talk_one_addnoise_further_dereverb}
\end{align}
where, different from $\widetilde{\hat{\mathbf{Z}}}(c,t,f)$ defined in Eq. (\ref{loss_MC_close_talk_one}), $\overline{\hat{\mathbf{Z}}}(d,t,f)=[\hat{Z}(d,t-I,f),\dots,\hat{Z}(d,t-\Delta,f)]^\T \in \CC^{I-\Delta+1}$ stacks $I-\Delta+1$ T-F units that are at least $\Delta$ ($>0$) frames in the past, and $\hat{\mathbf{h}}_{d}(f)\in \CC^{I-\Delta+1}$ is computed in the same way as Eq. (\ref{fcp_proj_mixture}).
$\hat{\mathbf{h}}_{d}(f)^{\H}\ \overline{\hat{\mathbf{Z}}}(d,t,f)$ is designed to absorb (or explain) the late reverberation component inside the close-talk speech, thereby driving $\hat{Z}(d)$ towards an estimate with less reverberation.

In Eq. (\ref{loss_MC_close_talk_one_addnoise_further_dereverb}), $\hat{Z}$ is constrained to estimate a dereverberated close-talk speech.
When combining this technique with supervised training on simulated mixtures for semi-supervised training, we should modify the $\mathcal{L}_{\text{sup,speech}}^\text{CTRnet}$ loss defined in Eq. (\ref{loss_supervised_speech}), where $\hat{Z}$ is however trained to fit close-talk speech.
We just replace the speaker's close-talk speech in Eq. (\ref{loss_supervised_speech}) with direct-path signal, which can be readily simulated along with close-talk speech.
Note that when noise modeling is not used, the summation in Eq.~(\ref{loss_MC_close_talk_one_addnoise_further_dereverb}) 
runs from $c=1$ to $C$ (rather than $C+1$), reducing to a 
modification of (\ref{loss_MC_close_talk_one}) instead of (\ref{loss_MC_close_talk_one_addnoise}).

\subsection{Inference of CTRnet}\label{run_time_target_estimation}

At inference time, we use $\hat{Z}(c)$ as the estimate of the close-talk speech for each speaker $c$.
We just need to run feed-forwarding once to obtain all the estimates.
All the FCP filtering operations are not needed at inference time.

We emphasize that speaker-activity timestamps are only needed for model training and not needed for inference.

\section{Cross-Talk Reduction for Separation}\label{PuLSS_description}

This section describes how the estimated close-talk speech produced by CTRnet is leveraged to train far-field speech separation models.
In our framework shown in Fig.~\ref{system_overview_figure}, CTRnet is trained first as described in Section \ref{CTRnet_description}.
Once trained, it is applied to all real-recorded close-talk mixtures in the training set to produce close-talk speech estimates, which are then used as fixed pseudo-labels for training PuLSS. The CTRnet parameters are not updated during PuLSS training.

\begin{figure*}
  \begin{center}
  \includegraphics[width=11.25cm]{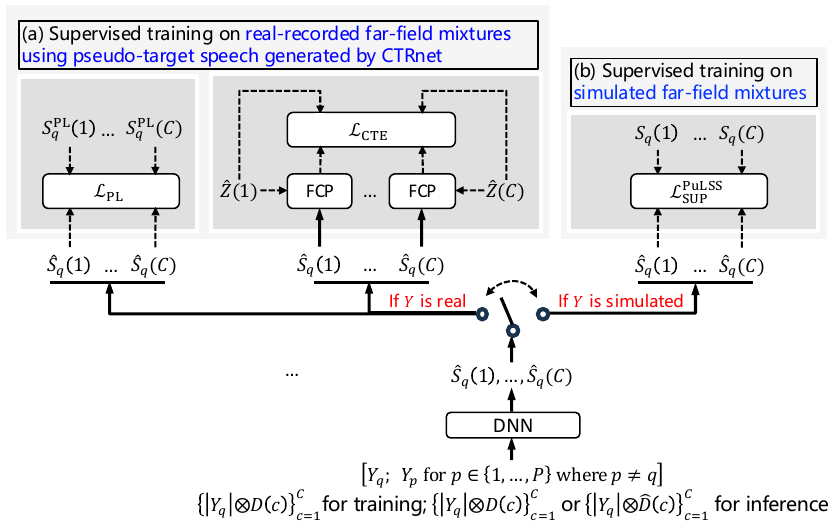}
  \end{center}
  \vspace{-0.3cm}
  \caption{
  Illustration of PuLSS. Best viewed in color.
  }\label{semi_supervised_PuLSS_figure}
  \vspace{-0.3cm}
\end{figure*}

This section describes PuLSS.
See Fig. \ref{semi_supervised_PuLSS_figure} for an illustration.

\subsection{Deriving Pseudo-Labels Based on Close-Talk Estimates}

With each close-talk estimate, we compute the speaker's image (ideally, direct-path signal) at a reference far-field microphone by estimating the RTF relating the speaker source signal to its image, and use the computed speaker image as the pseudo-target speech for the far-field mixture.
In detail, we first estimate the RTF $\hat{\mathbf{h}}_q(c,f) \in \CC^{L}$ using FCP \cite{Wang2021FCPjournal}:
\begin{align}\label{pseudo_direct_RIR}
\hat{\mathbf{h}}_q(c,f) = \underset{\mathbf{h}_q(c,f)}{\text{argmin}}
\sum\limits_t \frac{\Big| Y_q(t,f) - \mathbf{h}_q(c,f)^{\H}\ \Breve{\hat{\mathbf{Z}}}(c,t+\hat{K},f) \Big|^2}{\lambda_q(c,t,f)},
\end{align}
where $q\in\{1,\dots,P\}$ denotes a reference far-field microphone, $\Breve{\hat{\mathbf{Z}}}(c,t,f)=[\hat{Z}(c,t-L+1,f),\dots,\hat{Z}(c,t,f)]^\T \in \CC^{L}$ stacks a short window of $L$ T-F units in estimated close-talk speech, and $L$ is tuned to a small value (in this study $2$) to encourage $\mathbf{h}_q(c,f)$ to approximate the RTF of the direct-path signal.
In $\Breve{\hat{\mathbf{Z}}}(c,t+\hat{K},f)$, $\hat{K}$ is an estimated time delay (in frames) accounting for time-synchronization issues between close-talk and far-field microphones.
In our study, $\hat{K}$ is estimated via enumeration by solving the problem below:
\begin{align}\label{estimate_K}
\hat{K} = \underset{K\in \Psi}{\text{argmin}} \Big(\underset{\mathbf{h}_q(c,\cdot)}{\text{min}}
\sum\limits_{t,f} \frac{\Big| Y_q(t,f) - \mathbf{h}_q(c,f)^{\H}\ \Breve{\hat{\mathbf{Z}}}(c,t+K,f) \Big|^2}{\lambda_q(c,t,f)}\Big),
\end{align}
where
$\Psi=\{-E,\dots,0,\dots,E\}$
is a discrete set of hypothesized time delays (in frames), with $E$ (tuned to $9$ in this study) denoting the maximum hypothesized delay.
Since the filter tap is assumed small, the computation of Eq. (\ref{estimate_K}) is fast. 

With the estimated RTF, we compute the pseudo-label of speaker $c$ at the reference far-field microphone $q$
as follows:
\begin{align}\label{pseudo_direct_path}
S_q^{\text{PL}}(c,t,f) = \hat{\mathbf{h}}_q(c,f)^{\H}\ \Breve{\hat{\mathbf{Z}}}(c,t+\hat{K},f),
\end{align}
where the superscript ``PL'' means pseudo-label.

\subsection{Using Pseudo-Labels for Training Far-Field Models}\label{ssec:pulss_training}

With the pseudo-labels, we can train supervised models directly on far-field mixtures to realize far-field speech separation.
For speaker separation, we need to resolve the permutation problem \cite{Kolbak2017}. 
A common method is permutation invariant training (PIT) \cite{Kolbak2017}, but it only resolves permutations within each processing block.
For long-form audio, continuous speech separation \cite{chen2020continuous} processes overlapping blocks and stitches them along time, but it introduces a cross-block permutation problem that requires additional speaker reconciliation via, e.g., speaker embeddings and clustering \cite{raj2021integration}.
To avoid this complexity, we resolve the permutation problem by conditioning the separation model on speaker-activity timestamps.
This is conceptually similar to how GSS \cite{boeddeker2018front} leverages speaker-activity timestamps to guide its spatial clustering, albeit here within an end-to-end DNN framework.

We assume that, at training time, the timestamps of all speakers (i.e., $\{D(c)\}_{c=1}^C$) are available, and use them to compute masked mixture magnitude spectrograms $\{D(c)\otimes |Y_q|\}_{c=1}^C$, with $\otimes$ denoting point-wise multiplication, as additional input features (see Fig.~\ref{semi_supervised_PuLSS_figure}).
These $C$ masked spectrograms are concatenated with the mixture real and imaginary (RI) components and fed to the DNN, which produces $C$ separated outputs $\{\hat{S}_q(c)\}_{c=1}^C$ in a single forward pass.
This assigns each output channel to a specific speaker, removing permutation ambiguity across blocks.
Based on $\mathcal{G}(\cdot,\cdot)$ defined in Eq. (\ref{loss_G}), the loss function is defined as follows:
\begin{align}\label{PL_loss}
\mathcal{L}_{\text{PL}} = \sum_{c=1}^{C} \frac{ \sum_{t,f} \mathcal{G} \Big( \hat{S}_q(c,t,f), S_q^{\text{PL}}(c,t,f) \Big) }{ \sum_{t,f}| Y_{q}(t,f)|^\alpha }.
\end{align}

Although at training time we use oracle speaker-activity timestamps to compute input features, at inference time we can use timestamps estimated by a speaker diarization model to compute the features.
See later Section \ref{inference_PuLSS_description} for the details.

\subsection{Directly using Close-Talk Estimates for Training}

The pseudo-labels $S_q^{\text{PL}}$ produced via Eq. (\ref{pseudo_direct_path}), due to the linear filtering, often have a lower quality than $\hat{Z}$,
possibly limiting the performance of far-field separation.
To deal with this, besides using the $\mathcal{L}_{\text{PL}}$ loss in Eq. (\ref{PL_loss}) for training, we linearly filter the DNN estimates to approximate close-talk speech estimated by CTRnet and compute an additional loss:
\begin{align}\label{CTE_loss}
\mathcal{L}_{\text{CTE}} = \sum_{c=1}^C \frac{\sum_{t,f}\mathcal{G}\Big( \hat{Z}(c,t+\hat{K},f), \hat{\mathbf{o}}(c,f)^\H \grave{\hat{\mathbf{S}}}_q(c,t,f) \Big)}{\sum_{t,f} |Y_{d(=c)}(t,f)|^\alpha},
\end{align}
where $\hat{Z}(c)$ denotes the estimated close-talk speech of speaker $c$, $\hat{K}$ is computed via Eq. (\ref{estimate_K}) to account for time-synchronization issues, and $\grave{\hat{\mathbf{S}}}_q(c,t,f)=[\hat{S}_q(c,t-A,f),\dots,\hat{S}_q(c,t,f),\dots,\hat{S}_q(c,t+A,f)]^\T \in \CC^{A+1+A}$ stacks a short window of $A+1+A$ T-F units.
The linear filter $\hat{\mathbf{o}}(c,f) \in \CC^{A+1+A}$ is computed by FCP as follows:
\begin{align}\label{pseudo_reverse_RIR}
\hat{\mathbf{o}}(c,f) = \underset{\mathbf{o}(c,f)}{\text{argmin}}
\sum\limits_t \frac{\Big| \hat{Z}(c,t+\hat{K},f)-\mathbf{o}(c,f)^{\H}\ \grave{\hat{\mathbf{S}}}_q(c,t,f) \Big|^2}{\hat{\eta}(c,t+\hat{K},f)},
\end{align}
where $\hat{\eta}(c)$ is defined by replacing $Y_m$ in (\ref{FCPweight_ct}) with $\hat{Z}(c)$.

We combine the above loss functions for model training:
\begin{align}\label{total_PuLSS_real_loss}
\mathcal{L}_{\text{PL+CTE}}  = \mathcal{L}_{\text{PL}} + \delta \times \mathcal{L}_{\text{CTE}},
\end{align}
where $\delta\in \RR_{>0}$ is a tunable weighting term.
Note that if $\mathcal{L}_{\text{CTE}}$ is used alone for training, the DNN estimates would have a random gain level, as the linear filter in Eq. (\ref{CTE_loss}) can compensate for any gain levels in the DNN estimate.
We hence need $\mathcal{L}_{\text{PL}}$, which can penalize inaccurate gain estimation.

\subsection{Training PuLSS on Simulated and Real Mixtures}\label{ssec:pulss_real_simu}

Similarly to CTRnet, we can also train PuLSS by additionally including simulated mixtures.
If the input mixture is real-recorded, we use pseudo-label as the training target, while, if it is simulated, we use clean speech as the training target.
The loss for simulated mixtures can be defined as
\begin{align}\label{PuLSS_loss_simu}
\mathcal{L}_{\text{sup}}^\text{PuLSS} = \sum_{c=1}^{C} \frac{ \sum_{t,f} \mathcal{G} \Big( \hat{S}_q(c,t,f), S_q(c,t,f) \Big) }{ \sum_{t,f}| Y_{q}(t,f)|^\alpha },
\end{align}
where $S_q(c)$ is the direct-path signal of speaker $c$ at the reference microphone $q$.
The overall loss is defined, similarly to Eq. (\ref{CTRnet_loss_simu_real}) and by using a tunable weighting term
$\kappa_2$,
as
\begin{equation}
\mathcal{L}_\text{sup,PL+CTE}^\text{PuLSS} = \left\{
\begin{aligned}
&\kappa_2 \times \mathcal{L}_{\text{sup}}^\text{PuLSS}, \text{if input mixture is simulated}; \\
&\mathcal{L}_{\text{PL+CTE}}, \text{if input mixture is real-recorded}.
\end{aligned}
\right.
\label{PuLSS_loss_simu_real}
\end{equation}

\subsection{Inference of PuLSS}\label{inference_PuLSS_description}

At inference time, depending on the application scenarios, we have two setups.
The first one assumes that oracle speaker-activity timestamps are available (i.e., oracle speaker diarization) and fed to the trained model to predict target speech.
The second one estimates the speaker-activity timestamps via a speaker diarization model, and feeds the estimated timestamps directly to the trained model to predict target speech.

At inference time, we use $\hat{S}_q(c)$ as the separated speech for each speaker $c$.
We run feed-forwarding once to obtain all the $C$ estimates.
All the FCP filtering operations are not needed.

\section{Experimental Setup}\label{exp_setup_descriptin}

Our experiments aim at verifying whether CTRnet can accurately estimate close-talk speech and whether the estimated close-talk speech can serve as a good pseudo-label for real mixtures and help us develop better far-field speech separation models.
This section first introduces the CHiME-6 dataset \cite{Watanabe2020CHiME6} designed for conversational speech separation and recognition.
Next, we describe our data simulation procedure for supervised training.
We then provide details on the evaluation metrics, miscellaneous configurations, and baseline systems.

\subsection{CHiME-6 Dataset}\label{chime6_description}

The CHiME-6 dataset \cite{Watanabe2020CHiME6} consists of real-recorded conversational sessions, each captured in a different house.
Each session has $4$ speakers talking spontaneously for $120$--$150$ minutes.
Each speaker wears a binaural close-talk microphone to record their close-talk speech.
In some segments, close-talk recordings are missing due to the speaker removing the microphone or hardware malfunction.
The close-talk mixtures contain severe cross-talk, since the speakers are typically close to one another while talking.
Far-field mixture signals are captured by $6$ Kinect devices (each with $4$ microphones) distributed across the living room, kitchen, and dining room ($2$ devices per room), with speakers free to move between rooms.
Realistic noises that are typical in dinner-party scenarios are recorded at the same time along with speech. The task is to recognize each speaker's speech from the far-field mixtures and to output $4$ transcriptions (one per speaker), which requires accurate speaker diarization.
The sampling rate is $16$ kHz.

Our choice of CHiME-6 is dictated by the fact that it is a notoriously-difficult benchmark, primarily because its real-recorded signals are highly representative of the issues a deployed system must handle, such as microphone synchronization errors, 
signal clipping, frame dropping, microphone failures, moving speakers, time-varying speaker overlap ratios, and realistic environmental noises.
Among the scenarios featured in the recent CHiME-\{7,8\} DASR challenges \cite{cornell23_chime, Cornell2024CHiME8Description, Cornell2025CHiME78Analysis}, CHiME-6 is the most challenging.
More broadly, unlike datasets in office-meeting scenarios such as NOTSOFAR-1 \cite{vinnikovnotsofar}, AMI \cite{McCowan2006} and AliMeeting \cite{Yu2022M2MeT} which feature more structured conversations in acoustically controlled environments, CHiME-6 captures fully unconstrained dinner-party speech in real domestic settings, making it a perfect example of conversational speech \textit{in the wild}.
As
noted in Section \ref{ssec:frontend_separation_related}, the most successful speech separation algorithm on this dataset to date remains GSS \cite{boeddeker2018front}, a signal-processing method, with all the top teams in the CHiME-\{7,8\} challenges adopting GSS as their only speech separation module \cite{Wang2023USTCCHiME7,Ye2023IACASCHiME7,cornell23_chime, Cornell2025CHiME78Analysis}.

The official CHiME-6 dataset \cite{Watanabe2020CHiME6} consists of $16/2/2$ sessions ($\sim$$34/2/5$ hours) respectively in its training, validation and test sets.
We adopt this session partition but, to make our results directly comparable with the challenge submissions to the CHiME-\{7,8\} DASR challenges \cite{cornell23_chime,Cornell2024CHiME8Description}, we omit from training the two sessions that CHiME-7 DASR reassigned to its test set, resulting in only $14$ training sessions.
In our setup, the validation and test sets remain unchanged from CHiME-6 and same as the CHiME-8 DASR challenge, allowing direct comparison with all challenge submissions across the CHiME-\{6,7,8\} DASR challenges.

\subsection{Dealing with Long-Form Mixtures}\label{block_split_description}

In CHiME-6, as mentioned each session is long, lasting $120$ to $150$ minutes.
This is typical in conversational datasets.
We simply cannot feed-forward the entire signal of each session for training and inference.

For the training of CTRnet and PuLSS, we cut each session to $12$-second blocks with $11$-second overlap between consecutive blocks (i.e., extracting a $12$-second block every $1$ second), resulting in $123,339$ blocks ($\sim$$411$ hours) for model training.

\begin{figure}
  \begin{center}
  \includegraphics[width=\columnwidth]{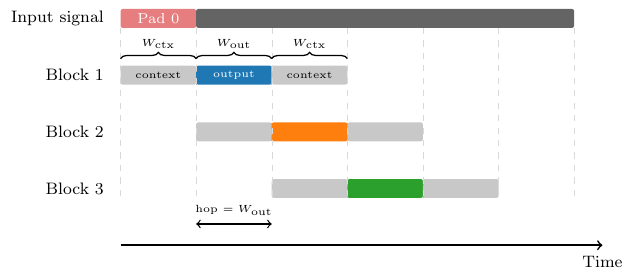}
  \end{center}
  \vspace{-0.3cm}
  \caption{Illustration of block-wise inference.
}
  \label{blockwise_inference_figure}
  \vspace{-0.3cm}
\end{figure}

For the inference of CTRnet and PuLSS, we apply the trained models block-wise to process each session, and stitch the processing results along time.
See Fig.~\ref{blockwise_inference_figure} for an illustration.
Each block has a total length of $W = W_{\text{ctx}} + W_{\text{out}} + W_{\text{ctx}}$, 
consisting of $W_{\text{ctx}}$ of context on each side, and $W_{\text{out}}$ of 
center output.
In this study, we set $W_{\text{ctx}}$ and $W_{\text{out}}$ to $4$ seconds, resulting in $W = 12$-second blocks. 
Only the DNN predictions in the center $W_{\text{out}}$ seconds are 
retained.
Consecutive blocks are shifted by $W_{\text{out}}$.

\subsection{Dealing with Binaural Close-Talk Microphones in CTRnet}

The close-talk mixtures in CHiME-6 are recorded by binaural microphones, meaning that each speaker has two close-talk microphones rather than one.
To address this, we investigate two strategies.
\textbf{Binaural Strategy \#1} considers the right-ear microphones as close-talk microphone while the left-ear one as far-field, resulting in $P+C$ far-field microphones and still $C$ close-talk microphones.
\textbf{Binaural Strategy \#2} averages the left- and right-ear mixtures and considers their average as the close-talk mixture, leading to $C$ close-talk and $P$ far-field mixtures.
The rationale is that the wearer's mouth is approximately in
the front direction of the two binaural microphones, and simply averaging the two channels is akin to delay-and-sum beamforming,
which can boost the SNR of close-talk speech, potentially improving cross-talk reduction.

\subsection{Dealing with Distributed Far-Field Microphone Arrays}

In CHiME-6, the devices used for recording far-field mixture signals are multiple Kinect devices, placed in a distributed manner in a random device geometry (but each Kinect device has the same, fixed microphone geometry).
In this case, we need to modify our algorithms for training CTRnet and PuLSS.

For CTRnet, we can use all the far-field microphones to compute the MC loss for model training.
This could lead to better cross-talk reduction, as more mixture constraints, afforded by the distributed microphone arrays, can be used for training.
The FCP filters can be configured long to cover the relative time delays of each speaker at different microphones.

For PuLSS, we can stack all the $24$ far-field microphone signals as input to train a DNN to predict the pseudo-target speech at a 
reference microphone of all the $24$ microphones.
We tried this idea but did not succeed, possibly because
in CHiME-6 different speakers exhibit significantly different time delays and energy levels at different arrays.
We hence use an alternative, where we only use the signals recorded by each microphone array as input (i.e., $4$-channel) to train a DNN to predict the pseudo-target speech at a designated reference microphone of the array.
At run time, we apply the DNN to process the entire session recorded by each array.
After that, we estimate the SNR of each speech segment (identified by oracle or estimated speaker-activity timestamps) of each speaker at each array based on the predicted signal and the mixture in the identified segment, and select the predicted signal with the highest
estimated SNR as the estimate of that speaker in the identified segment.

\subsection{Training Segment Sampling Based on Overlap Ratio}

For real-recorded conversations, speaker overlap ratio varies with time, as people tend to stay silent while the others are speaking and it is not common to have many speakers talking at the same time.
See Fig. \ref{sparse_overlap_figure} for an illustration.
This implicitly creates a data imbalance issue for training models on real data, as there are many more training blocks with low speaker overlap ratio than high overlap.
The resulting trained models could have limited capability at separating mixtures with high speaker overlap ratio.

To deal with this, we design a weighted sampling strategy, where the more speaker overlap ratio a training block has, the more likely it is selected for model training. The weight for training block $i$ is defined as follows:
\begin{align}\label{weighted_sampling}
w^{(i)} = 1 + \theta\times \frac{1}{T}\sum_{t=1}^T\Big(\max\big(1, \sum_{c=1}^C D^{(i)}(c,t)\big) - 1\Big),
\end{align}
where $\sum_{c=1}^C D^{(i)}(c,t)$ is the number of active speakers at frame $t$, $T$ the number of frames in the block, and $\theta \in \RR_{\geq0}$ a tunable hyper-parameter.
When there is no speaker overlap, $w^{(i)}=1$, and when every frame has $C$ speakers, $w^{(i)}=1+\theta\times (C-1)$, where $\theta$ controls the sampling weight.

\subsection{Miscellaneous Configurations for CTRnet and PuLSS}

TF-GridNet \cite{Wang2023TFGridNet} is used as the DNN architecture for CTRnet and PuLSS, as it has shown strong performance in many speech separation benchmarks.
We adopt two configurations, denoted as \textbf{V1} and \textbf{V2}, specified using the notation of \cite{Wang2023TFGridNet}.
The V1 model sets $D=100$, $B=4$, $I=2$, $J=2$, $H=200$, $L=4$ and $E=8$, while V2 changes $D=128$, $B=6$, $I=1$ and $J=1$.
The V1 model uses $\sim$$1/3$ the computation of V2, supporting fast experimentation for ablation studies. Note that the symbols
above refer to TF-GridNet hyper-parameters and are distinct from the identically-named variables in Table~\ref{summary_hyperparam}.

Both CTRnet and PuLSS are trained via complex spectral mapping \cite{Wang2020css}, which predicts the real and imaginary (RI) components of target signals based on the RI components of input mixture signals.
For PuLSS, we include the mixture magnitude masked by speaker-activity timestamps as additional input features (see Fig. \ref{semi_supervised_PuLSS_figure}) to avoid PIT as explained in Section~\ref{ssec:pulss_training}.

For STFT, the window and hop sizes are respectively set to $16$ and $8$ ms for CTRnet, and to $32$ and $16$ ms for PuLSS.
The square root of Hann window is used as the analysis window.

We use Adam
as the optimizer to train PuLSS and CTRnet.
Each epoch randomly samples $5\%$ of the training blocks.
When both simulated and real blocks are used for training, they have the same probabilities of being sampled.
The mini-batch size is $2$.
The learning rate starts from $10^{-3}$ and is halved if the validation loss is not improved in $2$ epochs.
We stop training when the learning rate is reduced to $6.25\times 10^{-5}$.
The $L_2$ norm for gradient clipping is set to $1.0$.

All the hyper-parameters listed in Table \ref{summary_hyperparam} are tuned based on the CHiME-6 validation set.

\subsection{Data Simulation for Supervised Training}\label{data_simulation_description}

CTRnet and PuLSS can be trained by using simulated mixtures in addition to real-recorded mixtures.
This subsection describes the method for data simulation.

For each real-recorded segment extracted from the CHiME-6 sessions for model training, we simulate a signal which is also $12$-second long.
It is synthesized by following the speaker-overlap patterns of the real-recorded segment.
In detail, given a $12$-second speaker-activity timestamp of each speaker (i.e., a zero-one vector), we first identify time ranges where the values are all ones (indicating active speech).
Next, we sample a speaker from clean speech databases including LibriSpeech \cite{PanayotoV2015} and EARS \cite{Richter2024EARS}, and place, in each of the identified time ranges, a sampled active speech segment of the speaker (that is shorter than the identified time range).
The active speech segments of each speaker are pre-identified by using the Pyannote
voice activity detection model\footnote{See \url{https://huggingface.co/pyannote/voice-activity-detection}.}, and pre-normalized to a sample variance of $1.0$.
We then scale each of the four $12$-second clean speech signals such that their energy level is sampled from the range $[-9, 9]$ dB.
Based on the \textit{Pyroomacoustics} toolkit,
the clean speech signals and up to $4$ noises sources sampled from the FSD50K dataset \cite{Fonseca2020FSD50k} and the pure noise signals in the CHiME-6 training data are placed in different locations in a simulated room (with reverberation time sampled from the range $[0.2, 0.7]$ s) to simulate noisy-reverberant far-field mixtures captured by a $4$-channel far-field microphone array, which is simulated based on the microphone-array geometry of the Kinect device.
The speech and noise signals are scaled such that the energy level between the summation of direct-path speech signals and that of reverberant noise signals  equals a value sampled from the range $[-20,20]$ dB.
Besides simulating far-field mixtures, for each speaker, we place a close-talk microphone to the speaker at a distance sampled from the range $[0.2,0.5]$ m to simulate close-talk mixture.
For both far-field and close-talk mixtures, a weak air-conditioning noise sampled from the REVERB dataset \cite{Kinoshita2016} is added.
In total, there are $123,339$ $12$-second blocks ($\sim$$411$ hours) simulated to train CTRnet and PuLSS.

\subsection{Evaluation Metrics}\label{eval_metrics_description}

The separated signal of each speaker, obtained via block-wise inference shown in Fig. \ref{blockwise_inference_figure}, has the same length as the input session.
Based on oracle or estimated speaker-activity timestamps, it is split into short utterances and then decoded by an ASR model.
We evaluate performance using two metrics from the CHiME-\{7,8\} DASR challenges \cite{cornell23_chime,
Cornell2024CHiME8Description, Cornell2025CHiME78Analysis}, which are suitable for multi-talker ASR evaluation.
When oracle diarization is used,
we report the concatenated minimum permutation word error rate (cpWER) \cite{Watanabe2020CHiME6} following the challenge protocol.
This enables direct comparison with all submissions to the CHiME-7 challenge, which has a track that assumes oracle diarization \cite{Cornell2025CHiME78Analysis}. 
When estimated diarization is used,
we report time-constrained cpWER (tcpWER) \cite{von2025word}, which extends
cpWER by time-aligning hypothesis words to reference speaker segments before scoring, so that segmentation errors also incur in insertion or deletion penalties.
We follow the scoring protocol (including the text normalization) of the CHiME-\{7,8\} DASR challenges summary paper \cite{Cornell2025CHiME78Analysis}.
This allows a direct comparison of our results with previous challenge submissions.

\subsection{ASR Models}\label{asr_models_description}

We evaluate our systems using two ASR models with different capability.
The first one, denoted as \textbf{Default}, is the ASR baseline provided by the CHiME-\{7,8\} challenge organizers \cite{cornell23_chime}\footnote{See \url{https://huggingface.co/popcornell/chime7_task1_asr1_baseline}.}.
It is an end-to-end encoder-decoder transformer model with hybrid CTC/attention based on a WavLM encoder.
Its training data includes the close-talk and far-field mixtures of CHiME-6, and far-field mixtures enhanced by GSS.
See Section 5.3 of \cite{cornell23_chime} for more details.
Note that the default ASR model favors GSS, as it utilizes GSS-enhanced signals for training.
This matters for our comparisons, since off-the-shelf ASR models trained without exposure to a separation frontend's output distribution often fail to benefit from it \cite{iwamoto2022bad,masuyama2026end}.
The second one, denoted as \textbf{Fine-tuned Parakeet}, is based on the Parakeet-TDT-0.6B-v3 model \cite{sekoyan2025canary}, a $600$M-parameter FastConformer model with a token-and-duration transducer decoder, pre-trained on ${\sim}1.7$ million hours of speech data.
We choose Parakeet-v3, as it is a representative state-of-the-art open English ASR model.
Its large-scale pre-training provides a strong starting point that is largely complementary to the dinner-party acoustic conditions of CHiME-6.
We fine-tune it on a combination of three data sources from the CHiME-6 training set: PuLSS-enhanced far-field mixtures, CTRnet-enhanced close-talk mixtures, and the original close-talk mixtures.
For all the three sources, oracle diarization is used to obtain segment boundaries.
Fine-tuning is performed using the NeMo toolkit \cite{kuchaiev2019nemo}, with full fine-tuning of the encoder, decoder, joint network, and duration head.
We use the AdamW optimizer with a learning rate of $5\times 10^{-5}$, a cosine annealing schedule with $5{,}000$ warm-up steps, and a weight decay of $10^{-3}$.
We train for $20$ epochs and select the best checkpoint based on the validation WER.
Hyper-parameters are tuned based on the CHiME-6 validation set using PuLSS-enhanced far-field signals with oracle diarization as input.

\subsection{Baseline Systems}\label{baseline}

We mainly consider two baseline approaches.
One is supervised speech separation \cite{WDLreview}, where DNN models are trained on simulated data. Based on the data simulated in Section \ref{data_simulation_description}, our supervised models are trained by penalizing the DNN estimates only using the $\mathcal{L}_{\text{sup,speech}}^{\text{CTRnet}}$ loss for CTRnet and $\mathcal{L}_{\text{sup}}^{\text{PuLSS}}$ for PuLSS.
The other one is GSS \cite{boeddeker2018front}, a signal processing method.
We use the GSS implementation provided in the CHiME-7 DASR challenge\footnote{See the training recipe at \href{https://github.com/espnet/espnet/blob/master/egs2/chime7_task1/asr1/local/run_gss.sh}{https://github.com/espnet/espnet/blob/master/egs2/ chime7\_task1/asr1/local/run\_gss.sh}}.
It first performs multi-channel speech dereverberation using the weighted prediction error algorithm,
and then computes a T-F mask-based minimum variance distortionless response beamformer for speech separation by using posterior T-F masks estimated by a spatial clustering module guided by speaker-activity timestamps provided by oracle or estimated speaker diarization \cite{boeddeker2018front}.
We emphasize that, so far, almost all the systems in public ASR challenges leverage GSS as the only module for speech separation \cite{Cornell2025CHiME78Analysis}.
Meanwhile, as CHiME-6 is a public dataset used in the CHiME-\{6,7,8\} challenges, which attracted broad participation, we can readily compare the results of our systems with many existing ones.

\begin{table*}[]
\footnotesize
\centering
\sisetup{table-format=2.2,round-mode=places,round-precision=2,table-number-alignment = center,detect-weight=true,detect-inline-weight=math}
\caption{\textsc{ASR Results of CTRnet on CHiME-6 Close-Talk Mixtures (\textbf{Diarization: Oracle}; ASR Model: Default; DNN: V1).}}\label{CTRnet_oracle_diar}
\setlength{\tabcolsep}{2pt}
\begin{tabular}{
l %
c %
c %
c %
S[table-format=2.1,round-precision=1] %
S[table-format=2.1,round-precision=1] %
c %
c %
S[table-format=2.0,round-precision=0] %
S[table-format=2.0,round-precision=0] %
c %
S[table-format=3.1,round-precision=1] 
S[table-format=3.1,round-precision=1] 
}
\toprule
 & & & & & & & & & & & \multicolumn{2}{c}{cpWER (\%)$\downarrow$} \\
\cmidrule(lr{9pt}){12-13}
ID & Systems & \makecell{Binaural\\Strategy} & {\makecell{\#Far-field\\mics ($P$)}} & {\makecell{Mag. compress.\\factor ($\alpha$)}} & {\makecell{Weight for\\$\mathcal{L}_\text{SA}$ ($\beta$)}} & \makecell{FCP\\denomin.} & \makecell{\#DNN\\estimates} & {\makecell{Reverb\\modeling ($\Delta$)}} & {\makecell{Sampling\\($\theta$)}} & \makecell{Noise\\modeling} & \multicolumn{1}{c}{Val.} & \multicolumn{1}{c}{Test} \\
\midrule
0 & Unprocessed mixture & {\#1} & {--} & {--} & {--} & {--} & {--} & {--} & {--} & {--} & 28.42 & 29.3902 \\
\midrule
1a & GSS ($4$-channel) \cite{boeddeker2018front} & {\#1} & {--} & {--} & {--} & {--} & {--} & {--} & {--} & {--} & 30.3918 & 32.6024 \\
1b & GSS ($8$-channel) \cite{boeddeker2018front} & {--} & {--} & {--} & {--} & {--} & {--} & {--} & {--} & {--} & 26.2241 & 28.2457 \\
\midrule
2 & Supervised CTRnet & {\#1} & {--} & 1.0 & {--} & {--} & 4 & {--} & {--} & {--} & 30.3595 & 37.8931 \\
\midrule
3a & Unsupervised CTRnet & {\#1} & $0\times 4+4$ & 1.0 & {--} & (\ref{FCPweight_ct}) & $4$ & {--} & {--} & {--} & 79.9664 & 76.9253 \\
3b & Unsupervised CTRnet & {\#1} & $1\times 4+4$ & 1.0 & {--} & (\ref{FCPweight_ct}) & $4$ & {--} & {--} & {--} & 81.9534 & 79.6804 \\
3c & Unsupervised CTRnet & {\#1} & $6\times 4+4$ & 1.0 & {--} & (\ref{FCPweight_ct}) & $4$ & {--} & {--} & {--} & 21.8305 & 25.5868 \\
\midrule
4a & Weakly-supervised CTRnet & {\#1} & $0\times 4+4$ & 1.0 & 1.0 & (\ref{FCPweight_ct}) & $4$ & {--} & {--} & {--} & 21.3363 & 24.8268 \\
4b & Weakly-supervised CTRnet & {\#1} & $1\times 4+4$ & 1.0 & 1.0 & (\ref{FCPweight_ct}) & $4$ & {--} & {--} & {--} & 21.3634 & 24.6001 \\
4c & Weakly-supervised CTRnet & {\#1} & $6\times 4+4$ & 1.0 & 1.0 & (\ref{FCPweight_ct}) & $4$ & {--} & {--} & {--} & 21.7914 & 24.99 \\
\midrule
5 & Weakly-supervised CTRnet & {\#2} & $6\times 4$ & 1.0 & 1.0 & (\ref{FCPweight_ct}) & $4$ & {--} & {--} & {--} & 20.4803 & 23.3069  \\
\midrule
6a & Semi-supervised CTRnet & {\#2} & $6\times 4$ & 1.0 & 1.0 & (\ref{FCPweight_ct}) & $4$ & {--} & {--} & {--} & 20.0472 & 22.4798 \\
6b & Semi-supervised CTRnet & {\#2} & $6\times 4$ & 1.0 & 0.1 & (\ref{FCPweight_ct}) & $4$ & {--} & {--} & {--} & 19.8774 & 22.2277 \\
6c & Semi-supervised CTRnet & {\#2} & $6\times 4$ & 0.3 & 0.1 & (\ref{FCPweight_ct}) & $4$ & {--} & {--} & {--} & 19.8417 & 22.3764 \\
6d & Semi-supervised CTRnet & {\#2} & $6\times 4$ & 0.3 & 0.1 & (\ref{FCPweight_ct_2}) & $4$ & {--} & {--} & {--} & 19.643 & 22.0009 \\
\midrule
7 & Semi-supervised CTRnet & {\#2} & $6\times 4$ & 0.3 & 0.1 & (\ref{FCPweight_ct_2}) & $4$ & 3 & {--} & {--} & \bfseries 19.4919 & 22.0354 \\
\midrule
8 & Semi-supervised CTRnet & {\#2} & $6\times 4$ & 0.3 & 0.1 & (\ref{FCPweight_ct_2}) & $4$ & {--} & 20 & {--} & \bfseries 19.4528 & 21.9084 \\
\midrule
9 & Semi-supervised CTRnet & {\#2} & $6\times 4$ & 0.3 & 0.1 & (\ref{FCPweight_ct_2}) & $4$ & 3 & 20 & {--} & \bfseries 19.5241 & \bfseries 21.8323 \\
\midrule

10a & Semi-supervised CTRnet & {\#2} & $6\times 4$ & 0.3 & 0.1 & (\ref{FCPweight_ct_2}) & $4+4$ & 3 & 20 & {(\ref{noise_averaging})} & 19.5751 & 22.0608 \\
10b & Semi-supervised CTRnet & {\#2} & $6\times 4$ & 0.3 & 0.1 & (\ref{FCPweight_ct_2}) & $4+4$ & 3 & 20 & {(\ref{noise_random})} & 19.5785 & 21.8722 \\

\bottomrule
\end{tabular}
\vspace{-0.3cm}
\end{table*}

\section{Evaluation Results on Close-Talk Mixtures}\label{evaluation_results_CTRnet_description}

Table \ref{CTRnet_oracle_diar} reports the evaluation results of CTRnet.
The hyper-parameters for FCP filtering are tuned to  $\xi=0.01$, and $I=13$ and $J=1$ (resulting in $15$-tap filters).
The weighting term used in $\mathcal{L}_{\text{MC+SA}}$ in Eq. (\ref{loss_MC+SA}) is tuned to $\beta=1.0$.
Since there are $4$ speakers in each session, we set $C=4$, meaning that the number of input microphones to CTRnet is also $4$.
The number of far-field microphones, $P$, is equal to $24$ ($=6\times 4$) in default, as there are $6$ Kinect devices, each with $4$ microphones, and all of them are used to compute the MC loss in default.

\subsection{Results of GSS and Supervised CTRnet}

The results of unprocessed mixtures are shown in row $0$, where we use the right-ear close-talk mixture (i.e., Binaural Strategy \#1) as the separation result for each speaker.
The cpWER on the test set is $29.4\%$.

In row $1$a, we use the right-ear close-talk mixtures to perform GSS, and in $1$b, we use both the left- and right-ear close-talk mixture for GSS.
Since there are $4$ speakers, we have $4$ input channels for $1$a and $8$ ($=2\times 4$) for $1$b.
We observe that $8$-channel GSS outperforms $4$-channel GSS, but the improvement over unprocessed mixtures is small (i.e., from $29.4\%$ to $28.2\%$ cpWER).
The small improvement could be because each speaker has very different SNRs at different close-talk microphones and in this case the target T-F masks at different close-talk microphones are significantly different.

Row $2$ reports the results of supervised CTRnet.
Although the simulated training data covers a wide range of acoustic conditions, the trained model performs much worse than the unprocessed mixture ($37.9$\% vs. $29.4$\% cpWER).
This is consistent with prior observations that supervised neural separation models trained on simulated data introduce artifacts and distortions that hurt downstream ASR when the backend has not been adapted to their output distribution \cite{Zhang2021ClosingGap, iwamoto2022bad, masuyama2026end}.

\subsection{Results of Un- and Weakly-Supervised CTRnet}

Row $3$a-$3$c report the results of unsupervised CTRnet, which uses the right-ear close-talk mixtures as input and is trained solely on real-recorded mixtures.
In $3$a, the left-ear close-talk microphones are considered as the far-field microphones for loss computation, and therefore $P=4$.
In $3$b, we additionally include the first far-field microphone array for loss computation, resulting in $P=1\times 4 + 4$ far-field microphone signals for loss computation.
In $3$c, we include all the $6$ far-field microphone arrays for loss computation, leading to $P=6\times 4 + 4$ channels.
We observe that the performance of unsupervised CTRnet heavily depends on the number of far-field microphones used for loss computation.
In $3$a and $3$b, unsupervised CTRnet does not work, while in $3$c, it works.

Row $4$a-$4$c report the results of weakly-supervised CTRnet, which is trained solely on real-recorded mixtures.
Comparing them with $3$a-$3$c, we observe that, even if $P$ is small (e.g., $P=4$ in $4$a), weakly-supervised CTRnet still works, and outperforms unsupervised CTRnet.

In row $5$, we switch to Binaural Strategy $\#2$, where we average each binaural close-talk mixture, and use the $4$ averaged signal as the input to CTRnet.
In this case, $P$ equals $6\times 4$.
This simple change leads to clear improvement over $4$c (from $25.0\%$ to $23.3\%$ cpWER).

\begin{table*}[]
\footnotesize
\centering
\sisetup{table-format=2.2,round-mode=places,round-precision=2,table-number-alignment = center,detect-weight=true,detect-inline-weight=math}
\caption{\textsc{ASR Results of PuLSS on CHiME-6 Far-Field Mixtures (\textbf{Diarization: Oracle}).}}\label{PuLSS_results_oracle_diar_default_ASR}
\vspace{-0.1cm}
\setlength{\tabcolsep}{7pt}
\begin{tabular}{
l %
c %
r %
S[table-format=2.0,round-precision=0] %
r %
c %
c %
S[table-format=2.1,round-precision=1] %
S[table-format=2.1,round-precision=1] %
}
\toprule
 & & & & & & & \multicolumn{2}{c}{cpWER (\%)$\downarrow$} \\
\cmidrule(lr{9pt}){8-9}
ID & System & {Loss function} & {\makecell{$\theta$}} & Pseudo-label & DNN & ASR backend & \multicolumn{1}{c}{Val.} & \multicolumn{1}{c}{Test} \\
\midrule
0 & Mixture & {--} & {--} & {--} & {--} & Default & 61.3543 & 62.6238 \\
\midrule
1 & \makecell{GSS (24-channel) \cite{boeddeker2018front}} & {--} & {--} & {--} & {--} & Default & 32.4128 & 38.5425 \\
\midrule
2 & Supervised & $\mathcal{L}_\text{sup}^\text{PuLSS}$ in (\ref{PuLSS_loss_simu}) & {--} & ID $9$ of Table \ref{CTRnet_oracle_diar} & V1 & Default & 42.5621 & 49.0369 \\
\midrule
3a & PuLSS & $\mathcal{L}_{\text{PL}}$ in (\ref{PL_loss}) & {--} & ID $9$ of Table \ref{CTRnet_oracle_diar} & V1 & Default & 31.4295 & 35.3647 \\
3b & PuLSS & $\mathcal{L}_{\text{PL+CTE}}$ in (\ref{total_PuLSS_real_loss}) & {--} & ID $9$ of Table \ref{CTRnet_oracle_diar} & V1 & Default & 28.9227 & 32.2179 \\
\midrule
4a & PuLSS & $\mathcal{L}_{\text{sup,PL}}^\text{PuLSS}$ in (\ref{PuLSS_loss_simu_real}) & {--} & ID $9$ of Table \ref{CTRnet_oracle_diar} & V1 & Default & 29.3015 & 33.6725 \\
4b & PuLSS & $\mathcal{L}_\text{sup,PL+CTE}^\text{PuLSS}$ in (\ref{PuLSS_loss_simu_real}) & {--} & ID $9$ of Table \ref{CTRnet_oracle_diar} & V1 & Default & 27.5861 & 31.3019 \\
\midrule
5 & PuLSS & $\mathcal{L}_\text{sup,PL+CTE}^\text{PuLSS}$ in (\ref{PuLSS_loss_simu_real}) & 20 & ID $9$ of Table \ref{CTRnet_oracle_diar} & V1 & Default & 27.2838 & 31.0426 \\
\midrule
6 & PuLSS & $\mathcal{L}_\text{sup,PL+CTE}^\text{PuLSS}$ in (\ref{PuLSS_loss_simu_real}) & 20 & ID $10$b of Table \ref{CTRnet_oracle_diar} & V1 & Default & 27.1735 & 30.912 \\
\midrule
7a & PuLSS & $\mathcal{L}_\text{sup,PL+CTE}^\text{PuLSS}$ in (\ref{PuLSS_loss_simu_real}) & 20 & ID $10$b of Table \ref{CTRnet_oracle_diar} & V2 & Default & 26.6487 & 29.9652 \\
7b & PuLSS & $\mathcal{L}_\text{sup,PL+CTE}^\text{PuLSS}$ in (\ref{PuLSS_loss_simu_real}) & 20 & ID $10$b of Table \ref{CTRnet_oracle_diar} & V2 & Fine-tuned Parakeet & \bfseries 16.7 & \bfseries 19.5 \\
\bottomrule
\end{tabular}
\vspace{-0.3cm}
\end{table*}

\subsection{Results of Semi-Supervised CTRnet}

Row $6$a improves weakly-supervised CTRnet (in row $5$), which is trained solely on real-recorded mixtures, by including supervised training on simulated mixtures, leading to semi-supervised CTRnet.
We set $\kappa_1$ in Eq. (\ref{CTRnet_loss_simu_real}) to $1.0$.
This change improves the performance from $23.3\%$ in row $5$ to $22.5\%$ cpWER in $6$a.
This improvement indicates the effectiveness of including supervised learning on simulated mixtures.

In row $6$b, we tune $\beta$ from $1.0$ to $0.1$; in $6$c, we tune $\alpha$ from $1.0$ to $0.3$; and in $6$d, we change the FCP denominator from Eq. (\ref{FCPweight_ct}) to (\ref{FCPweight_ct_2}).
Each change produces slight improvement on the validation set (i.e., from $20.0\%$ in $6$a to $19.9\%$ in $6$b, to $19.8\%$ in $6$c, and to $19.6\%$ in $6$d).
The three changes combined produce clearly better cpWER on the test set (i.e., $22.0\%$ cpWER in $6$d vs. $22.5\%$ cpWER in $6$a).

Row $7$ reports the results of modeling reverberation in close-talk speech.
We tune the prediction delay $\Delta$ based on the set of $\{1,2,3,4\}$.
When it is tuned to $3$,
on the validation set we obtain slightly better results over $6$d (i.e., $19.5\%$ vs. $19.6\%$ cpWER), which does not dereverberate close-talk speech.

Row $8$ reports the results of using training segment sampling, where the sampling weight $\theta$ in Eq. (\ref{weighted_sampling}) is tuned based on the set of $\{5, 10, 20, 40, 80\}$.
Slightly better cpWER is observed when it is set to $20$ (i.e., $21.9\%$ in row $8$ vs. $22.0\%$ in $6$d on the test set).

Row $9$ combines reverb modeling and training segment sampling, yielding better results at $21.8\%$ on the test set.

Row $10$a and $10$b report the results of including noise modeling in semi-supervised CTRnet, where the DNN is trained to additionally output $4$ noise estimates.
Using Eq. (\ref{noise_random}) to combine the noises estimates works better than (\ref{noise_averaging}).
Although informal listening tests suggest that the additional noise output can absorb some of the ambient noise, incorporating noise modeling does not improve cpWER.
Nonetheless, as we will report later in Table \ref{PuLSS_results_oracle_diar_default_ASR}, CTRnet trained with noise modeling leads to a slightly better PuLSS model.

\subsection{Discussion}

In Table \ref{CTRnet_oracle_diar}, we use the default ASR model, which is trained
on unprocessed close-talk and far-field mixtures, and GSS-enhanced signals (see Section \ref{asr_models_description}), so both unprocessed mixtures and GSS outputs are in-domain at inference, while CTRnet outputs are not.
In addition, with oracle diarization providing per-speaker segment boundaries, the ASR can learn to implicitly attend to the \textit{centered} speaker within each segment, which partly accounts for the strong performance of unprocessed mixtures and raises the bar for any separation frontend to be useful.
Our weakly- and semi-supervised CTRnet variants outperform GSS (in row $1$a and $1$b) and unprocessed mixtures (in row $0$), indicating their effectiveness.
In addition, they outperform supervised CTRnet (in row $2$), suggesting the benefits of training on real-recorded data.

\section{Evaluation Results on Far-Field Mixtures}\label{evaluation_results_PuLSS_description}

This section reports the evaluation results of PuLSS on far-field mixtures.
$\delta$ in Eq. (\ref{total_PuLSS_real_loss}) is  tuned to $20$, and $\kappa_2$ in (\ref{PuLSS_loss_simu_real}) is set to $1.0$.
The number of input channels to DNN is $4$, which equals the number of microphones in each Kinect device.

\subsection{Results of GSS, Supervised, and PuLSS Approaches}

Table \ref{PuLSS_results_oracle_diar_default_ASR} reports the results on far-field mixtures.
In row $0$, the cpWER of unprocessed mixtures is $62.6\%$, which is quite large, indicating the difficulty of this task.
It is obtained by directly using the first microphone of the first far-field array for ASR.
For GSS, we use all the far-field microphone signals as input, which is $24$-channel ($=6\times 4$).
For the \textit{Supervised} model, we use the supervised training setup in Fig. \ref{semi_supervised_PuLSS_figure}(b) for training, with all the other setup same as PuLSS.
From rows $1$ and $2$, we observe that supervised PuLSS improves over the unprocessed mixture ($49.0\%$ vs.\ $62.6\%$ cpWER) but performs substantially worse than GSS ($38.5\%$ cpWER).

In comparison, for PuLSS models trained on the real-recorded mixtures using the pseudo-labels derived from the CTRnet in row $9$ of Table \ref{CTRnet_oracle_diar}, the results are clearly better.
When using the $\mathcal{L}_{\text{PL}}$ loss in Eq. (\ref{PL_loss}), we obtain $35.4\%$ cpWER in $3$a.
When using $\mathcal{L}_{\text{PL+CTE}}$ in~(\ref{total_PuLSS_real_loss}) with $\delta$ tuned to $20$, we improve the cpWER to $32.2\%$ in $3$b.
Further including supervised learning on simulated mixtures based on the $\mathcal{L}_\text{sup,PL+CTE}^\text{PuLSS}$ loss in~(\ref{PuLSS_loss_simu_real}) improves cpWER to $31.3\%$ in $4$b.
In row $5$, further performing training segment sampling improves the performance from $31.3\%$ to $31.0\%$ cpWER.
In row $6$, we switch to the CTRnet in row $10$b of Table~\ref{CTRnet_oracle_diar}, which performs noise modeling, to compute the pseudo-labels for training PuLSS.
Slightly better cpWER is observed.

In row $7$a, we switch the configuration of TF-GridNet from V1 to V2, which uses more computation.
This improves PuLSS from $30.9\%$ to $30.0\%$ cpWER.
The experiments so far are based on the default ASR model provided with the CHiME-7 DASR challenge.
In $7$b, we fine-tune the pre-trained Parakeet ASR model on the separated close-talk mixtures from CTRnet, separated far-field mixtures from PuLSS, and the original close-talk mixtures.
Dramatic improvement is observed (from $30.0\%$ to $19.5\%$ cpWER).
This gain comes from the ASR change alone, reflecting the inherently stronger pre-trained Parakeet backbone and its adaptation to the output distribution of our separation models.
We use this ASR model in default in subsequent sections.
For fair comparison, the GSS results reported in subsequent sections are obtained with the same Parakeet model fine-tuned identically as PuLSS but on GSS-enhanced signals instead, so that both PuLSS and GSS benefit from a matched degree of ASR adaptation.

\subsection{Comparison with Existing Systems - Oracle Diarization}\label{comparison_with_existing_studies_description_oracle_diar}

Table \ref{PuLSS_comparison_with_others_oracle_diar} compares the performance of PuLSS with existing approaches, all using oracle diarization at run time.
Again, the results for GSS are obtained by fine-tuning the Parakeet-v3 model in the same way as for PuLSS, but using GSS-enhanced signals instead of PuLSS-enhanced ones.
PuLSS achieves strong ASR performance at $19.5\%$ cpWER, slightly improving over the previous best ($19.8\%$ cpWER) obtained by the USTC system \cite{Wang2023USTCCHiME7}, which uses an ensemble of multiple ASR models and iterative multi-stage decoding, with the first-pass ASR output used to refine speaker-activity timestamps for GSS.
On the other hand, it significantly outperforms GSS ($19.5\%$ vs. $29.7\%$ cpWER on the test set).

From the gray rows in Table \ref{PuLSS_comparison_with_others_estimated_diar}, we observe that, when using oracle diarization, applying CTRnet can reduce the cpWER of close-talk mixtures from $19.5\%$ to $15.0\%$.
This result indicates the effectiveness of CTRnet, and represents a performance upper bound for the downstream PuLSS models.

Note that the comparison between PuLSS and GSS isolates frontend quality and is a fair comparison, while the comparison of PuLSS with previous CHiME-\{7,8\} challenge submissions in Table \ref{PuLSS_comparison_with_others_oracle_diar} (and later Table \ref{PuLSS_comparison_with_others_estimated_diar}) are not similarly matched, as each challenge submission uses its own ASR backend, which is usually an ensemble of multiple models (see also \cite{Cornell2025CHiME78Analysis}) and is much more complicated than ours.
The challenge-submission comparison is included to reflect overall system performance.

\begin{table}[t]
\footnotesize
\centering
\sisetup{table-format=2.2,round-mode=places,round-precision=2,table-number-alignment = center,detect-weight=true,detect-inline-weight=math}
\caption{\textsc{ASR Results on CHiME-6 Far-Field Mixtures\\(\textbf{Diarization: Oracle}; ASR Model: Fine-Tuned Parakeet).}}
\label{PuLSS_comparison_with_others_oracle_diar}
\vspace{-0.1cm}
\setlength{\tabcolsep}{8pt}
\begin{tabular}{
l %
c %
S[table-format=2.1,round-precision=1] 
S[table-format=2.1,round-precision=1]
}
\toprule
& & \multicolumn{2}{c}{cpWER (\%)$\downarrow$} \\
\cmidrule(lr{9pt}){3-4}
System & Challenge & \multicolumn{1}{c}{Val.} & \multicolumn{1}{c}{Test} \\

\midrule
ESPnet baseline \cite{cornell23_chime} & CHiME-7 & 32.4 & 35.5 \\
NVIDIA NeMo \cite{cornell23_chime} & CHiME-7 &  21.6 & 25.7 \\
BUT-FIT \cite{Karafiat2023BUTCHiME7} & CHiME-7 & 23.8 & 27.6 \\
NPU \cite{Mu2023NPUCHiME7} & CHiME-7 & 24.9 & 29.6 \\
U. of Cambridge \cite{Deng2023UofCambridgeCHiME7} & CHiME-7 & 22.0 & 26.2 \\
USTC \cite{Wang2023USTCCHiME7} & CHiME-7 & 19.8  & 19.8 \\
IACAS-Thinkit \cite{Ye2023IACASCHiME7} & CHiME-7 & \bfseries 15.4 & 23.9  \\
NTT\cite{Kamo2024NTTCHiME8} & CHiME-8 & 19.8 & 24.0 \\
STCON \cite{MitrofanoV2024STCONCHiME8} & CHiME-8 & 18.5  & 23.0 \\
\midrule
GSS ($24$-channel) \cite{boeddeker2018front} & {--} & 24.8 & 29.7 \\
\midrule
PuLSS & {--} & 16.7 & \bfseries 19.5 \\
\midrule
\rowcolor{shadecolor}
Close-Talk Mixtures & {--} & 18.7 & 19.5 \\
\rowcolor{shadecolor}
\quad + CTRnet & {--} & 11.6 & 15.0 \\
\bottomrule
\multicolumn{4}{l}{
\textit{Note}: Systems using close-talk mixtures as input are marked in gray.}
\end{tabular}

\vspace{0.3cm}

\footnotesize
\centering
\sisetup{table-format=2.2,round-mode=places,round-precision=2,table-number-alignment = center,detect-weight=true,detect-inline-weight=math}
\caption{\textsc{ASR Results on CHiME-6 Far-Field Mixtures\\(\textbf{Diarization: Estimated}; ASR Model: Fine-Tuned Parakeet).}}
\label{PuLSS_comparison_with_others_estimated_diar}
\vspace{-0.1cm}
\setlength{\tabcolsep}{8pt}
\begin{tabular}{
l %
c %
S[table-format=2.1,round-precision=1] 
S[table-format=2.1,round-precision=1] 
}
\toprule
& & \multicolumn{2}{c}{tcpWER (\%)$\downarrow$} \\
\cmidrule(lr{9pt}){3-4}
System & Challenge & \multicolumn{1}{c}{Val.} & \multicolumn{1}{c}{Test} \\

\midrule
ESPnet baseline \cite{cornell23_chime} & CHiME-7 & 65.7 & 85.2 \\
NVIDIA NeMo \cite{cornell23_chime} & CHiME-7 & 45.9 & 63.8 \\
BUT-FIT \cite{Karafiat2023BUTCHiME7} & CHiME-7 & 61.4 & 77.6 \\
NPU \cite{Mu2023NPUCHiME7} & CHiME-7 & 57.4 & 76.9 \\
U. of Cambridge \cite{Deng2023UofCambridgeCHiME7} & CHiME-7 & 44.5 & 55.4 \\
USTC \cite{Wang2023USTCCHiME7} & CHiME-7 & 35.7 & 44.8 \\
IACAS-Thinkit \cite{Ye2023IACASCHiME7} & CHiME-7 & 30.5 & 33.5 \\
NTT\cite{Kamo2024NTTCHiME8} & CHiME-8 & 25.5 & 35.3 \\
STCON \cite{MitrofanoV2024STCONCHiME8} & CHiME-8 & \bfseries 22.8 & 33.6 \\
\midrule
GSS ($24$-channel) \cite{boeddeker2018front}  & & & \\
\quad + STCON diar. \cite{MitrofanoV2024STCONCHiME8} & {--} & 30.1 & 37.9  \\
\quad + USTC diar. \cite{Wang2023USTCCHiME7} & {--} &  29.4  &  33.5  \\
\midrule
PuLSS & & & \\
\quad + STCON diar. \cite{MitrofanoV2024STCONCHiME8} & {--} & 24.4 & 31.7 \\
\quad + USTC diar. \cite{Wang2023USTCCHiME7} & {--} & 26.4 & \bfseries 28.5 \\
\midrule
\rowcolor{shadecolor}
Close-Talk Mixtures & & & \\
\rowcolor{shadecolor}
\quad + STCON diar. \cite{MitrofanoV2024STCONCHiME8} & {--} & 28.1 & 35.3 \\
\rowcolor{shadecolor}
\quad + USTC diar. \cite{Wang2023USTCCHiME7} & {--} & 25.2 & 27.2 \\
\bottomrule
\multicolumn{4}{l}{
\textit{Note}: Systems using close-talk mixtures as input are marked in gray.
}
\end{tabular}
\vspace{-0.3cm}
\end{table}

\subsection{Comparison with Existing Systems - Estimated Diarization}\label{comparison_with_existing_studies_description_estimated_diar}

Table \ref{PuLSS_comparison_with_others_estimated_diar} reports results with estimated diarization at run time.
To assess PuLSS's robustness to diarization errors, we evaluate it using diarization outputs from two existing systems with markedly different speaker-boundary quality on CHiME-6: (a) the USTC system \cite{Wang2023USTCCHiME7}, which achieves the best Jaccard error rate (JER) \cite{Ryant2019DIHARD2} on CHiME-6 ($28.0\%$ on the test set)\footnote{JER figures were obtained from \cite{Cornell2025CHiME78Analysis}.}, and (b) the STCON system \cite{MitrofanoV2024STCONCHiME8}, the overall winner of the CHiME-8 challenge but with substantially weaker diarization on CHiME-6 (JER $38.6\%$ on the test set).
The latter therefore stress-tests the tolerance of PuLSS to speaker-activity boundaries with more errors.
For both diarization outputs, we replace the oracle speaker-activity timestamps with the estimated ones when computing the input features for PuLSS (see Fig. \ref{semi_supervised_PuLSS_figure}). PuLSS is trained only with oracle diarization and is not re-trained when using estimated speaker-activity timestamps at inference.
Even so, PuLSS achieves strong ASR performance with both diarization outputs, in both cases surpassing the previous best tcpWER on CHiME-6 (reported by IACAS-Thinkit \cite{Ye2023IACASCHiME7}) by a clear margin ($28.5\%$ vs. $33.5\%$ tcpWER).
These results show that PuLSS generalizes robustly across diarization quality and, as a front-end method, consistently outperforms GSS even with non-oracle diarization ($28.5\%$ vs. $33.5\%$ tcpWER for GSS when USTC diarization is used).

\section{Limitations}\label{limitation_description}

It should be noted that our method has several limitations.

Close-talk mixtures often contain non-verbal sounds of the wearer, such as chewing, breathing,
and laughing.
CTRnet
tends to preserve such sounds in its close-talk estimates, which can create difficulties for training PuLSS since the same non-verbal sounds are often too weak to be captured by far-field microphones.
A related issue is that such sounds may not be reliably annotated as speaker activity by speaker-diarization systems or even human annotators, introducing potential inconsistencies between the timestamps used to condition PuLSS and the actual content of the close-talk pseudo-labels.

Our framework assumes that the maximum number of speakers $C$ is fixed at $4$.
This is rarely a concern in practice, as the maximum number of speakers active within a $12$-second processing segment is generally small in conversational scenarios.
The same assumption also underlies widely-used speaker diarization systems such as Pyannote \cite{bredin2020pyannote} and EEND-VC \cite{kinoshita2021integrating}.
On the other hand, there are existing supervised speech separation algorithms \cite{Lee2024Boosting} that can deal with a large, unknown number of speakers.
They can be adapted in a straightforward way to our setup.

A further consideration is the breadth of empirical evaluation.
Although CHiME-6 is so far the most challenging real-recorded conversational benchmark,
spanning unconstrained dinner-party conversations in real domestic environments with realistic noises, microphone synchronization issues, signal clipping, and moving speakers (see Section \ref{chime6_description}), it remains a single acoustic scenario.
This work focuses on it because its difficulty makes it the most informative single benchmark for stress-testing real-data training, and because it allows direct comparison with a large body of CHiME-\{7,8\} challenge submissions.
Extending CTRnet and PuLSS to other conversational domains such as
AMI \cite{McCowan2006} and AliMeeting \cite{Yu2022M2MeT}
is straightforward,
since the framework requires only paired close-talk and far-field recordings.

Finally, two potentially beneficial training variants are not explored: training PuLSS directly with estimated (rather than oracle) speaker-activity timestamps to better match inference-time conditions, and end-to-end joint fine-tuning of PuLSS together with the downstream ASR model.
We leave these as natural extensions for future work.

\section{Conclusions}

We have proposed CTRnet and PuLSS, a two-stage framework for far-field speech separation in real-recorded conversational scenarios.
By formulating cross-talk reduction as a blind deconvolution problem, CTRnet jointly estimates close-talk speech and the RTFs to their reverberant images, and can be trained directly on real-recorded pairs of close-talk and far-field mixtures.
The resulting close-talk estimates serve as effective pseudo-labels for training PuLSS, a supervised far-field speech separation model that operates on real-recorded multi-channel far-field mixtures.
On the challenging CHiME-6 dataset, PuLSS achieves state-of-the-art ASR performance under both oracle and estimated diarization, surpassing all previous CHiME-\{7,8\} challenge submissions as well as GSS (the de-facto current state-of-the-art frontend approach for real conversational data) while remaining robust across different diarization systems.
Moving forward, we plan to investigate end-to-end fine-tuning of PuLSS with downstream ASR models and improved modeling of distributed microphone arrays.

\bibliographystyle{IEEEtran}
\bibliography{references.bib}

\vspace{-33pt}

\begin{IEEEbiography}
[{\includegraphics[width=1in,height=1.25in,clip,keepaspectratio]{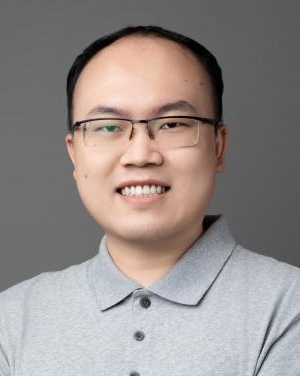}}]{Zhong-Qiu Wang}
received the B.E. degree in computer science and technology from Harbin Institute of Technology, Harbin, China, in $2013$, and the Ph.D. degree in computer science and engineering from The Ohio State University, Columbus, OH, USA, in $2020$.
He is currently a tenure-track associate professor in the Department of Computer Science and Engineering at Southern University of Science and Technology, Shenzhen, China.
He was a Postdoctoral Research Associate with Carnegie Mellon University, Pittsburgh, PA, USA, from $2021$ to $2024$, and a visiting research scientist at Mitsubishi Electric Research Laboratories, Cambridge, MA, USA, from $2020$ to $2021$.
His research interests include speech separation, robust automatic speech recognition, microphone array signal processing, and deep learning, aiming at solving the cocktail party problem.
He serves as an Action Editor for the Neural Networks journal, and was a committee member in Audio and Acoustic Signal Processing Technical Committee (AASP-TC).
\end{IEEEbiography}

\vspace{8pt}

\begin{IEEEbiography}
[{\includegraphics[width=1in,height=1.25in,clip,keepaspectratio]{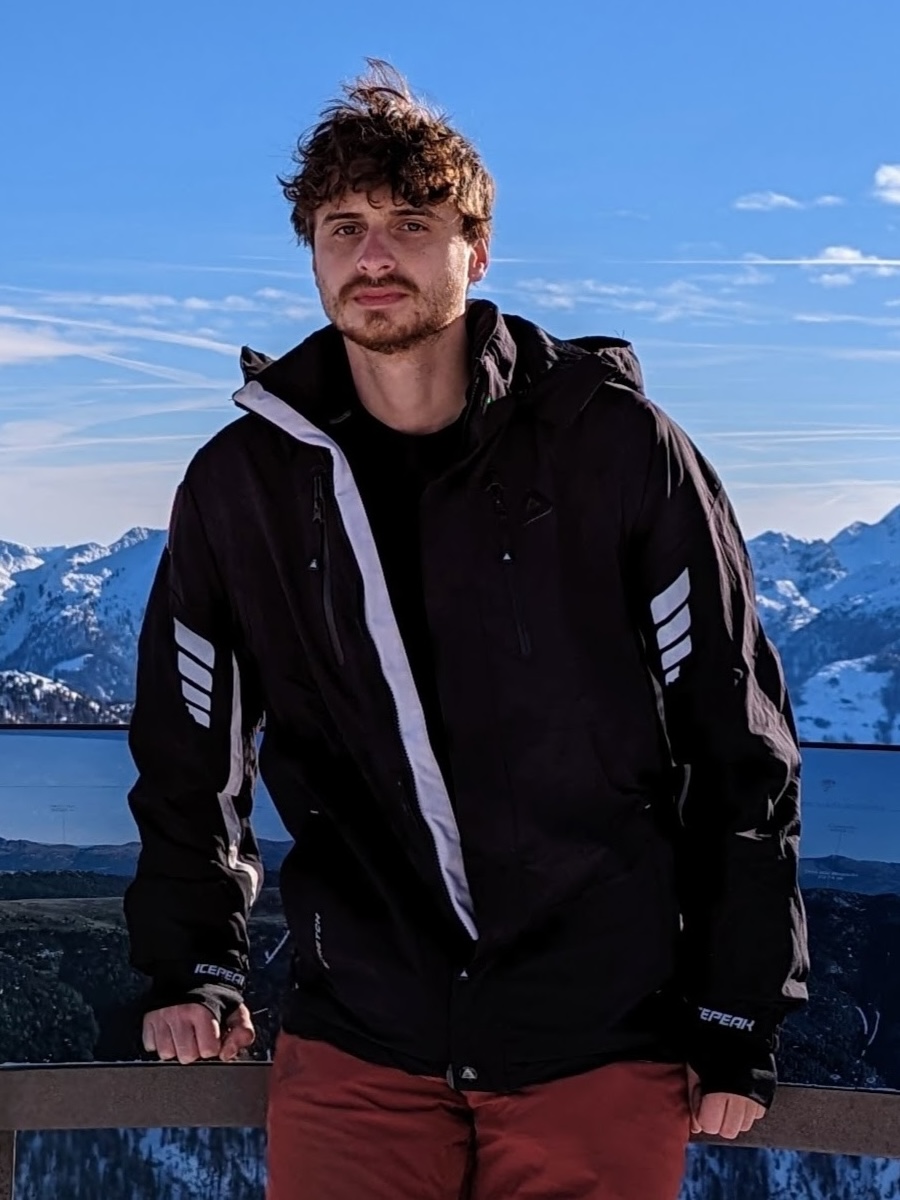}}]{Samuele Cornell} is currently a postdoctoral research associate at Carnegie Mellon University at the Language Technologies Institute within Prof. Shinji Watanabe research group (WAVLab). 
He got a Master degree in electronic engineering (summa cum laude) at Università Politecnica delle Marche in $2019$ and, in $2023$, at the same institution, a doctoral degree in Information Engineering.
His research interests are mainly in the area of robust speech processing (speech enhancement, speech separation, diarization, automatic speech recognition) for distant multi-talker conversational scenarios, and also in the broader field of machine listening (sound event detection and classification) with over $50$ publications in these fields. 
He is also author and has significant contributions in several popular open-source speech-processing toolkits (such as SpeechBrain, ESPNet, Asteroid source separation) and has organized and co-organized popular audio processing challenges in the fields of sound event detection, robust speech processing and speech enhancement such as DCASE Task 4 ($2022$, $2021$, $2024$),
CHiME-7/8 DASR, and URGENT ($2024$ and $2025$). 
\end{IEEEbiography}

\end{document}